\documentclass[preprint,aps,showpacs,prb]{revtex4}
\usepackage{amsfonts}
\usepackage{graphicx}
\usepackage{amsmath}
\usepackage{natbib}

\begin{document}
\title{sd-type Exchange Interactions in Nonhomogeneous  
Ferromagnets}

\author{W. N. G. Hitchon$^{1}$}
\author{G. J. Parker$^{2}$\footnote{Current address: GE Global Research,
One Research Circle, Niskayuna, New York 12309}}
-\author{A. Rebei$^{2}$ \footnote{Corresponding author:arebei@mailaps.org}}

\affiliation{
	$^{1}$Department of Electrical and Computer Engineering, University of Wisconsin-Madison, Madison, Wisconsin 53706, USA\\
$^{2}$Seagate Research Center, Pittsburgh, Pennsylvania 15222, USA
}

\date{04-20-04 }

\begin{abstract}
Motivated by a need to  understand spin-momentum transport in 
CPP (current
perpendicular to the plane) structures,
a quantum field theoretical treatment of spin-spin 
interactions in ferromagnets is presented. \ The 
sd interaction of 
the
conduction electrons and the magnetic medium is treated 
non-perturbatively from first 
principles in real
space. \ The 
localized magnetic
moments also interact with each other  through a Heisenberg exchange 
potential. \ To take 
into account correlation effects, a second quantized 
formulation is used. 
\ The semi-classical limit is taken by using a 
coherent-state 
path-integral
technique which also allows us to go beyond 
a linear-response approach. \ We derive a set of coupled 
equations of motion for the nonuniform
magnetization, the spin current and the 
two-point correlation functions of the
magnetization. \ The rate of change of the magnetization 
is shown to obey
a generalized Landau-Lifshitz equation that takes into 
account interaction
with the conduction electrons. \ Within the relaxation
time approximation 
it is  shown that 
 the polarization of the conduction electrons obeys 
a diffusion equation. \ The diffusion tensor, 
which has off-diagonal terms due to the sd exchange 
interaction,  is now  explicitly 
dependent on the magnetization of the medium. \ We also 
 show that 
the magnetization fluctuations satisfy a diffusion-type equation. \ The 
derived equations are used in two illustrative examples.
\end{abstract}

\maketitle

\newpage

\section{Introduction}

Spin-spin interactions in ferromagnetic metals are of paramount 
importance in
today's GMR recording heads. \ There is also currently great 
interest in the
magnetic recording industry in using spin currents, instead of magnetic
fields, to switch the magnetization in a writer device. \ In this case a
polarized electronic current is needed, such that the net spin of the
polarization becomes effectively another magnetic source which 
induces an 
 interaction with the magnetic moments of the medium. \ 
One widely used approximation is to separate the degrees 
of freedom of the current from those of the local magnetic 
moment. \ This latter separation is not justified in conducting 
metals but it nevertheless produced good results in some 
cases. \cite{Berger} \ This paper explores in detail the 
consequences for the spin accumulation problem in ferromagnets
of assuming that the interaction 
between the conduction electrons and the 
local moments is of the sd exchange 
type.
\ This interaction can give rise to 
what is now known as spin-momentum transfer (SMT) in 
magnetic multilayers. \ This latter 
mechanism has been predicted by
Berger \cite{Berger} and Slonczewski \cite{Slon} and later verified
experimentally by various groups. \cite{Cornell, Zurich} \ Other interaction
mechanisms between the conduction 
electrons  and the magnetization 
vector have been proposed
since the Berger-Slonczewski 
work. \cite{Heidi,Waintal,Zhang,Stiles} \ In 
previous work, the interaction of the 
polarized current with the magnetization has  not been
treated self-consistently. \ In fact the equations of motion 
were based on
those of a similar system, that of a current interacting with magnetic
impurities. \cite{Langreth} \ We believe that this approach 
is not suitable for transition metals.

\bigskip

\ In this work, we start from a 
microscopic
description of the conduction electrons 
 and the ferromagnetic medium and 
then take the
semi-classical limit to derive equations for macroscopic quantities 
of physical significance to experiment and other phenomenological 
approaches. \ Although the derivations are somewhat 
complex, one can go to the main results (e.g., Eqs. 
~\ref{current2} and  \ref{diff}  which 
are generalized Boltzmann-type equations) and 
see that the correct physics is contained in them.

\ Our results include those of 
reference \onlinecite{Zhang}
and in fact our treatment should provide a basis for the more 
phenomenological  work.  \ We do not assume 
that the
magnetization is uniform as in  previous work and we will focus 
mainly on the diffusive regime. 
\ We use many-body field theoretical methods to describe 
the system of
magnetic moments plus conduction electrons. \ Even though 
only a single particle picture 
is needed, the methods we
use permit us to treat the magnetic part of the problem
 and the conduction 
electrons
 on the same
formal level. \ This allows us to derive transport 
equations for the conduction electrons
and the local magnetic  moments and include relaxation effects without 
recourse to more
phenomenological modeling. \ Exchange effects, which are important in transition metals, are 
also included self-consistently.  \ Finite-temperature properties 
are naturally
included through the use of a 
path-integral formulation
of the problem.\cite{Chou} \ Including spin-dependent 
interactions in a
transport problem means that we have to deal with many indexes.
\ Path-integrals are helpful with book-keeping and hence 
simplify the
discussion considerably as compared to 
Ref. \onlinecite{Langreth}.\ Finally a
path-integral representation helps in making consistent 
approximations to the
derived transport equations.

\bigskip

\indent The paper is organized as follows. \ In Sec. II, 
we introduce the
Hamiltonian of the interacting electron-magnetization 
system in second
quantized form. \ In Sec. III, a functional $\mathbb{Z}$, which we call
the universal generating functional, is defined in terms of 
the density matrix
of the system and  external sources. \ This functional has a
path-integral representation and will generate all possible 
correlation
functions. \ For the local magnetic moments we adopt a coherent 
state representation which 
 is most suitable for a semi-classical treatment of 
the
medium such as presented here. \ In Sec. IV, we derive equations of
motion for the magnetization (a modified Landau-Lifshitz 
equation) and the spin accumulation. \ These latter equations are then 
used to derive equations for the 
  correlation 
functions of the magnetization. \  We also  show that non-uniform magnetization 
of the medium gives rise to a spin accumulation effect similar to that 
due to interfaces.\cite{Berger} \ This is one of the main results 
of this work.    \ In
Sec. V, we show how to 
solve these equations perturbatively and derive the terms 
that give rise to the spin
momentum transfer effect for a system with 
non-uniform magnetization. \ Finally, we derive a diffusion-type equation for 
the magnetization fluctuations where the diffusion coefficient is 
dependent on the exchange integral. \ In section VI, we show 
some applications of the derived spin accumulation equations in simple 
cases where the magnetization is nonuniform in the direction of flow of the 
current. \ 
\ In the
last section, we summarize our results.
\bigskip

%
%

\bigskip

\section{The Model}

\begin{figure}
[ptb]
\begin{center}
\includegraphics[
natheight=8.770900in,
natwidth=6.198100in,
height=2.0859in,
width=4.4255in
]
{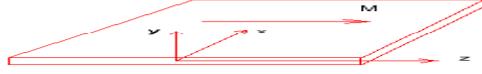}
\caption{Geometry of the magnetic sample. }
\label{slab}
\end{center}
\end{figure}

We start from a quantum field picture of the  
electronic current and the
magnetic moments of the localized electrons in a 
thin slab of thickness comparable to the 
mean free path, Fig. \ref{slab}. \ In the following 
we do not include explicitly 
an electric field, but we assume that it is part of the spin-independent 
part of the Hamiltonian. \ Its inclusion has been done in Ref. \onlinecite{hcr}. 
\ For the magnetic medium, a
Heisenberg Hamiltonian is assumed. \ We  explicitly take into 
account only
exchange effects among the 
Heisenberg spins. \ Other important 
effects such as the demagnetization 
field term can be
 added phenomenologically. \ The conduction electrons will
be assumed to be in equilibrium with the magnetic 
medium since they relax much
faster than the magnetic moments. \ This is the case in 
assisted spin-momentum
transfer switching. \ In fact the magnetic moments can be regarded as in
contact with a Fermi bath with spin degrees of freedom. \ In this case
switching may be regarded as a dissipative effect accompanied 
by a shift in the magnon energies.\cite{rebei4} \ Our work 
is a natural generalization of  the
model used by Langreth and Wilkins \cite{Langreth} to study 
spin resonance
in dilute magnetic alloys. \
The conduction electron field ${\Psi}_{s}$ satisfies 
the usual anti-commutation relations,%

\begin{eqnarray}
\left\{  \Psi_{s} (\mathbf{r}),\Psi_{s^{\prime}} (\mathbf{r}^{\prime
})\right\}   &  = & \left\{  \Psi_{s} ^{\dagger}(\mathbf{r}),\Psi_{s^{\prime}}
^{\dagger}(\mathbf{r}^{\prime})\right\}  = 0 , 
\end{eqnarray}
\begin{eqnarray}
\left\{  \Psi_{s} (\mathbf{r}),\Psi_{s^{\prime}} ^{\dagger}(\mathbf{r}
^{\prime})\right\}   &  = \delta_{ss^{\prime}} \delta\left(  \mathbf{r}
-\mathbf{r}^{\prime}\right)  ,\nonumber
\end{eqnarray}
where $s$ is a spin index $\left( s = \pm \frac{1}{2} \right)$.
\ The 
number density operator of the electrons  is given by
\begin{equation}
\rho_{ss} \left(  \mathbf{r}\right)  =\Psi_{s} ^{\dagger} (\mathbf{r})\Psi
_{s}(\mathbf{r}).
\end{equation}

\bigskip

\noindent In this system, the electrons are treated as 
non-interacting, i.e. no Coulomb interaction, and hence the
electron field can be expanded in terms of single 
particle wave-functions:
\begin{align}
\Psi_{s} (\mathbf{r})  &  =\sum_{i}\phi_{i} \left(  \mathbf{r}\right)  a_{s,i}
\; ,\\
\Psi_{s} ^{\dagger}(\mathbf{r})  &  =\sum_{i}\phi_{i}^{\ast} \left(
\mathbf{r}\right)  a_{s,i}^{\dagger} \; ,\nonumber
\end{align}
where $a_{s} (a_{s}^{\dagger})$ is the annihilation (creation) operator 
 for a particle of spin $s$. \
The total Hamiltonian of the system is composed of a conduction
 electron part, $H_{s}$,
a magnetic part, $H_{d}$ and an interaction part, $H_{sd}$,

\begin{equation}
H=H_{s}+H_{d}+H_{\mathbf{s}d} \; .
\end{equation}
The conduction electron part in a ferromagnet, such as Fe, is due to 4s-type electrons.
\ The magnetic part is however due to 3d-type electrons. \ For the 
 free conduction  electrons
 we have
\begin{eqnarray}
H_{s} 
&  = & \int\frac{\hbar^{2}}{2m}\nabla\Psi^{\dagger}\left(  \mathbf{r}\right)
\cdot\nabla\Psi\left(  \mathbf{r}\right)  d\mathbf{r}
\end{eqnarray}
where $\mathbf{\Psi}$ is the two-component Fermi field,
\begin{equation}
\Psi=\left(
\begin{array}
[c]{c}
\Psi_{1}\\
\Psi_{2}
\end{array}
\right) .
\end{equation}
The magnetic medium is microscopically a lattice with a spin vector
$\mathbf{S}_{i}$ at each lattice point $i$. \ Since we are interested in the
continuum limit of this model, we can define a macroscopic spin vector for the
medium
\begin{equation}
\mathbf{S}\left(  \mathbf{r}\right)  =\sum_{i=1}^{N}\mathbf{S}_{i}%
\delta\left(  \mathbf{r}-\mathbf{r}_{i}\right)  \; .
\end{equation}
The magnetization vector is then simply given by the volume 
average of the
global spin

\begin{align}
\mathbf{M}\left(  \mathbf{x}\right)   &  =g\mu_{B}\frac{\mathbf{S}\left(
\mathbf{x}\right)  }{V} \; ,
\end{align}
where $g$ is the spectroscopic splitting factor and $\mu_{B}$ is 
the Bohr 
magneton of the electron.
The spin vector has an $SU \left(  2 \right)  $ representation and
consequently so does the magnetization vector, e.g.
\begin{align}
\left[  M_{\mathbf{x}}\left(  \mathbf{x}\right)  ,M_{y}\left(  \mathbf{x}%
^{\prime}\right)  \right]  =2ig\mu_{B}M_{z}\left(  \mathbf{x}\right)
\delta\left(  \mathbf{x}-\mathbf{x}^{\prime}\right).
\end{align}
We employ units such that $g \mu_{B} = \hbar = 1$. \ Hence the magnetization will have 
{\it{opposite}} sign to that usually defined in the literature. \
The Hamiltonian of the spins is taken to be of the Heisenberg type,
\begin{equation}
H_{d}=-\frac{1}{2}\sum_{ij}J\left(  \mathbf{r}_{i}-\mathbf{r}_{j}\right)
\mathbf{S}_{i}\cdot\mathbf{S}_{j}- B\cdot\sum_{i}\mathbf{S}_{i}.
\end{equation}
We take account of exchange only and assume the spins to be in an 
external
field $\mathbf{B}$. In the continuum limit, we can write the exchange term in
an integral form,
\begin{equation}
\frac{1}{2}\sum_{i,j=1}^{N}\mathbf{S}_{i}\cdot\mathbf{S}_{j}J\left(
\mathbf{r}_{i}-\mathbf{r}_{j}\right)  =\frac{1}{2}\int d\mathbf{r}
d\mathbf{r}^{\prime}\sum_{\mu=1,2,3}\mathbf{S}_{\mu}\left( \mathbf{r}\right)
  J\left(  \mathbf{r}-\mathbf{r}^{\prime}\right)  \mathbf{S}_{\mu}\left(
\mathbf{r}^{\prime}\right)  .
\end{equation}
Contributions from the magnetostatic field can also 
be included. \ For the typical thin structures we are interested 
in, the magnetization of the medium is usually held in-plane by 
the magnetostatic  field. \ The energy term associated with this field 
is usually non-local but for a thin film 
it can be well approximated by 
a local term proportional to the normal component of the 
magnetization,
\begin{equation}
H_{demag} =  \frac{1}{2} K \int d \mathbf{x} \left[ \mathbf{S} \left( 
\mathbf{x} \right) \cdot \mathbf{n} 
\right]^{2},
\end{equation}
where $\mathbf{n}$ is a normal unit vector to the surface and 
$K$ is a constant depending on the shape of the slab. \ This 
latter term will lower the symmetry of the problem even further.  
\ For simplicity in the following we leave out this term and discuss it 
elsewhere. \cite{rebei4}

\bigskip

The interaction between the electrons and the localized spins is 
taken to be
of the s-d type, of the form
\begin{equation}
H_{sd}=-\frac{{\lambda}}{2}
\int d\mathbf{x}\left(  \mathbf{\Psi}^{\dagger}\left(  \mathbf{x}\right)
\mathbf{\overrightarrow {\mathbf{{\sigma}}}}\mathbf{\Psi}\left(  \mathbf{x}\right)  \right)  \cdot\mathbf{S}\left(
\mathbf{x}\right)
\end{equation}
where $\lambda$ is a coupling constant of the order of $0.1\; eV$ 
  and $\overrightarrow{\mathbf{\sigma}}$ is a vector whose
components are the Pauli matrices,
\begin{equation}
\left[  \sigma_{i}, \sigma_{j} \right]  = 2i\epsilon^{ijk} \sigma_{k} .
\end{equation}
$\epsilon^{ijk}$ is the antisymmetric unit tensor. \noindent The full
Hamiltonian is then the sum of all the above terms,

\begin{align}
H  &  =\sum_{\alpha=1,2}\int d\mathbf{r}\; \Psi_{\alpha}^{*}\left(
\mathbf{r}\right)  \left\{  \frac{1}{2m}\mathbf{p} ^{2}\left(  \mathbf{r}
\right)  + U\left(  \mathbf{r}\right)  -\frac{1}{2}\mathbf{\sigma}
\cdot\mathbf{B}\right\}  \Psi_{\alpha}\left(  \mathbf{r}\right) \\
&  -\frac{{\lambda}}{2}\int d\mathbf{r}\sum_{\alpha,\beta=1}^{2}\sum_{i=1}^{3}
\Psi_{\alpha}^{*}\left(  \mathbf{r}\right)  \sigma_{\alpha\beta}^{\left(
i\right)  }\Psi_{\beta}\left(  \mathbf{r}\right) {S}_{i}\left(
\mathbf{r}\right) \nonumber\\
&  -\frac{1}{2}\int d\mathbf{r}d\mathbf{r}^{\prime}\sum_{i=1}^{3}\left\{  \;
{S}^{i}\left(  \mathbf{r}\right)  J \left(  \mathbf{r}-\mathbf{r}
^{\prime}\right)  {S}_{i}\left(  \mathbf{r}^{\prime}\right)  +2\mu
_{s}{B}^{i}\left(  \mathbf{r}\right)  \delta\left(  \mathbf{r} -
\mathbf{r^{\prime}} \right)  {S}_{i }\left(  \mathbf{r^{\prime}
}\right)  \; \right\} \nonumber
\end{align}
where we have included a spin-independent external potential $U\left(
\mathbf{r} \right)$ for the 
conduction electrons. \ The external magnetic field $\mathbf{B}$ can be taken
to be time-dependent and/or spatially dependent in the following 
treatment.

\bigskip

\section{The Universal Generating Functional}

\bigskip

In this section, we introduce what we call a universal generating functional
from which we derive equations of motion for the magnetization and the spin
current. \ This functional is defined in terms of the density 
matrix $\rho$ of
the current-medium system and can be used to study equilibrium as well as
non-equilibrium properties.\cite{Schwinger} \ We are at present interested
only in the near-equilibrium case. \ A general overview of the 
method appeared in Ref. \onlinecite{Chou}. \ Application 
of this method to spin systems is however almost
absent. \ As far as we know, Refs. \onlinecite{Langreth} and 
\onlinecite{Manson} are 
the only work that apply non-equilibrium methods to spin systems.
 \ One of the other advantages of this method is that it allows
equal treatment of thermal effects and non-thermal effects. \ The extension
beyond a linear-response approach is also easily achieved, at 
least formally. This becomes essential when we are interested in 
questions that involve switching of the magnetization. \cite{rebei5}
 We will give a general outline of the method as we apply it 
to the particular s-d exchange system since the steps used in the 
derivation of the equations of motion are hard to find in one 
place. \ The importance of these methods, which are 
hardly used in magnetism, is potentially very great.

\bigskip

\ To motivate the structure of the functional we are 
about to introduce, we
recall the structure of the density matrix elements. \ In 
quantum mechanics,
we usually use the energy method to solve a problem. \ However, in transport
problems, time-dependent methods are essential. \ Hence the evolution of the
density matrix comes into play. \ One of the ways to calculate density matrix
elements is through a path-integral representation.\cite{Feynman} \ We have
recently used such a formalism to treat the problem of fluctuations and
dissipation in coherent 
magnetization.\cite{Rebei}$^{,}$\cite{rebei2} \ The density matrix
elements at a given time t are usually written in terms of those at an 
earlier
time using propagators running from past to present and from present
 to past,
\begin{eqnarray}
\left\langle \Phi\left|  \rho\left(  t\right)  \right|  \Phi^{\prime
}\right\rangle  &  = &\left\langle \Phi\left(  t\right)  \left|  \rho\right|
\Phi^{\prime}\left(  t\right)  \right\rangle \label{density}\\
&  = & \int\mathfrak{D}\Phi_{1}\mathfrak{D}\Phi_{2}\;
\left\langle \Phi\left(  t\right)
|\Phi_{1}\left(  t_{0}\right)  \right\rangle \left\langle \Phi_{1}\left(
t_{0}\right)  \left|  \rho\right|  \Phi_{2}\left(  t_{0}\right)  
\right\rangle
\left\langle \Phi_{2}\left(  t_{0}\right)  |\Phi^{\prime}\left(  t\right)
\right\rangle \nonumber\\
&  \equiv & \int\mathfrak{D}\Phi_{1}\mathfrak{D}\Phi_{2}\;\left\langle \Phi_{1}\left|
\rho\left(  t_{0}\right)  \right|  \Phi_{2}\right\rangle \mathfrak{G}^{\left(
+\right)  }\left(  \Phi\left(  t\right)  ,\Phi_{1}\left(  t_{0}\right)
\right)  \mathfrak{G}^{\left(  -\right)  }\left(  \Phi_{2}\left(  t_{0}\right)
,\Phi^{\prime}\left(  t\right)  \right) 
\end{eqnarray}

\bigskip

\noindent The propagators $\mathfrak{G}^{(+)}$ and
 $\mathfrak{G}^{(-)}$ are then
written in terms of path-integrals. This is equivalent to using the
Schr$\overset{..}{o}$edinger equation to solve for the evolution of the
density matrix. \ In a transport problem, we instead introduce a functional 
of
the density matrix. \ This functional is then made to depend on new virtual
sources $\eta_{1}$, $\eta_{1}^{\ast}$, $\eta_{2}$, $\eta_{2}^{\ast}$
, $\mathbf{J}_{1}$ and $ \mathbf{J}_{2}$. \ These sources are 
coupled to the conduction electrons' field and the magnetic
moments of the medium which will enable us to generate all kinds of
correlation functions and their time-evolution. \ The functional is then 
given
in terms of a trace formula,

\begin{equation}
{ {\mathbb{Z}}\left[  \;\eta_{1},\;\eta_{1}^{\ast},\;\eta_{2},\;\eta
_{2}^{\ast},\;\mathbf{J}_{1},\;\mathbf{J}_{2}, \; \rho\; \right]  \; \; =}
\end{equation}
\[
{ {Tr\;}}\left\{  \;{ \rho\left(  t_{0} \right)  \;}\left(
{ \mathcal{T}^{\;-1}\exp}\left[  { -i\int d\mathbf{x\;}}\left(
\;{ \eta_{2}^{\ast}\left(  \mathbf{x}\right) \cdot \mathbf{\Psi}\left(
\mathbf{x}\right)  +\mathbf{\Psi}^{+}\left(  \mathbf{x}\right) \cdot \eta
_{2}\left(  \mathbf{x}\right)  +\mathbf{J}_{2}\left(  \mathbf{x}\right)
\cdot\mathbf{S}\left(  \mathbf{x}\right)  \;}\right)  \right]  \right)
\right.
\]
\[
\;\;\ \ \ \ \ \ \ \ \ \ \;\times\left.  \left(  \mathcal{T}\exp\left[  \;i
\int
d\mathbf{x\;}\left(  \;{ \eta}_{1}^{\ast}\left(  \mathbf{x}\right) \cdot
\mathbf{ \Psi}\left(  \mathbf{x}\right)  +\mathbf{ \Psi}^{+}\left(
\mathbf{x}\right) \cdot { \eta}_{1}\left(  \mathbf{x}\right)  +\mathbf{ J}
_{1}\left(  \mathbf{x}\right)  \cdot \mathbf{ S}\left(  \mathbf{x}\right)
\;\right)  \;\right]  \right)  \;\right\}
\]
\bigskip

\noindent where $\mathcal{T}$ is the time-ordering operator. \ It orders
operators with the earliest time argument to the right. \ $\mathcal{T}^{-1}$
is the inverse of $\mathcal{T}$. \ The external sources $\eta_{1}$ and
$\eta_{2}$ are two-component classical (i.e. Grassmann) sources 
 where \ $\eta_{1}$ 
 $\left( \eta_{2} \right)$ 
and $\eta_{1}^{*}$  $ \left( \eta_{2}^{*}\right) $   are treated as 
independent. \ The operators are all 
written
in the Heisenberg representation. \ The need for both time-ordering operators
is clearly seen through the Feynman-Vernon Influence formalism, Eq.
$(\ref{density})$. \ The coefficients of the expansion of the functional
$\mathbb{Z}$ in terms of its arguments give all possible correlation
 functions
of the system. \ A few of the lowest order correlations are stated below.
\ They can be easily verified by differentiating the functional 
$\mathbb{Z}$ with respect to the virtual sources.
\ For example, to get the expectation values of the conduction 
electrons' field, we
differentiate $\mathbb{Z}$  with respect to $\eta_{1}$ or $\eta_{2}$,

\begin{equation}
\frac{1}{\mathbb{Z}}\frac{\delta\mathbb{Z}}{\delta\eta^{\ast}_{1 \lambda}
\left(  \mathbf{x}\right)  }| _{\eta=\eta^{\ast}=\mathbf{J}=0} =
i\langle\mathcal{T}{\Psi} _{\lambda}(\mathbf{x})\rangle\equiv
i\langle {\Psi} _{1\lambda}(\mathbf{x})\rangle\; ,
\end{equation}
\begin{equation}
\frac{1}{\mathbb{Z}}\frac{\delta\mathbb{Z}}{\delta\eta^{\ast}_{2 \lambda}
\left(  \mathbf{x}\right)  }| _{\eta=\eta^{\ast}=\mathbf{J}=0} =-i\langle
\mathcal{T}^{\; -1}{\Psi} _{\lambda}(\mathbf{x})\rangle\equiv
-i\langle {\Psi} _{2\lambda}(\mathbf{x})\rangle\; .
\end{equation}
\bigskip
\noindent Similarly to get expectation values for the magnetization, we
differentiate with respect to the external sources $\mathbf{J}_{1}$ and
$\mathbf{J}_{2}$,
\begin{equation}
\left.  \frac{1}{\mathbb{Z}}\frac{\delta\mathbb{Z}}{\delta{J}
_{1i}\left(  \mathbf{x}\right)  }\right|  _{\eta=\eta^{\ast}=\mathbf{J}=0} =
i\langle\mathcal{T} {S}_{i}(\mathbf{x})\rangle\equiv i\langle
{S}_{1i}(\mathbf{x})\rangle
\end{equation}
\begin{equation}
\left.  \frac{1}{\mathbb{Z}}\frac{\delta\mathbb{Z}}{\delta {J}
_{2i}\left(  \mathbf{x}\right)  }\right|  _{\eta=\eta^{\ast}=\mathbf{J}
=0}=-i\langle\mathcal{T}^{\; -1} {S}_{i}(\mathbf{x})\rangle
\equiv-i\langle{S}_{2i}(\mathbf{x})\rangle\; .
\end{equation}

\bigskip

\noindent Higher order correlations can be obtained simply via higher 
order differentiations.

\bigskip

The Hilbert space for the conduction  electrons  and 
the local magnetic moments of the medium is the product of
the corresponding Hilbert spaces,
\begin{equation}
\left|  \mathbf{\Phi},\mathbf{\Omega}\right\rangle \equiv\left|
\Phi\right\rangle \otimes\left|  \Omega\right\rangle ,
\end{equation}
where $\left|  \mathbf{\Phi}\right\rangle $ is a many-body fermionic state 
representing the conduction electrons and
$\left|  \mathbf{\Omega}\right\rangle $ is a magnetic moment state. \ The
magnetic moment states will be represented in terms of spin-coherent states
(SCS) \cite{Perm} (and references therein). \ Since the operators are
initially taken to be in the Heisenberg picture, then in the presence of the
additional external sources, $\eta_{1}$, $\eta_{2}$, and $\mathbf{J}$, the
states are no longer time independent. \ Now we write the functional formula
in terms of a path-integral. \ Since we are not interested in the transient
behavior of the interaction between the current and the magnetic moments, we
assume that the external electric field was turned on a long time ago and we
will eventually set $t_{0} = -\infty$. \ Reference \onlinecite{rebei4} treats the case 
where $t_0$ is kept finite in a finite size thin film.  \ Moreover, we assume that the density
matrix is separable initially, i.e.,
\begin{equation}
\rho\left(  t=-\infty\right)  =\rho_{s}\left(  t=-\infty\right)  
\otimes
\rho_{d}\left(  t=-\infty\right)  ,
\end{equation}
where $\rho_{s}$ is the density matrix of the conduction s-electrons and
$\rho_{d}$ is that of the local magnetic moments.
\bigskip
Now let $\left|  \mathbf{\Phi}_{i},\mathbf{\Omega}_{i}\right\rangle $ be an
initial over-complete set of states for the operators $\mathbf{\Psi} \left(
\mathbf{r}, t_{0} \right)  \otimes1 $ and $1\otimes\mathbf{S}\left(
\mathbf{r}, t_{0} \right)  $. \ Similarly, we let $\left|  \mathbf{\Phi}%
_{c},\mathbf{\Omega}_{c}\right\rangle $ be an over-complete set of states for
the operators $\mathbf{\Psi} \left(  \mathbf{r}, t_{c} \right)  \otimes1 $ and
$1\otimes\mathbf{S}\left(  \mathbf{r} , t_{c}\right)  $ at the time $t_{c}$.
\ At each intermediate time, we define similar states. \ Then the functional
$\mathbb{Z}$ can be written as follows
\[
{ {\mathbb{Z}}\left[  \;\eta_{1},\;\eta_{1}^{\ast},\;\eta_{2},\;\eta
_{2}^{\ast},\;\mathbf{J}_{1},\;\mathbf{J}_{2}, \; \rho\; \right]  \; \; = }
\]
\[
\int\mathfrak{D}\mathbf{\Phi}_{i}^{*}\mathfrak{D}\mathbf{\Phi}_{i} 
\mathfrak{D}
\mathbf{\Omega}_{i} \int\mathfrak{D}\mathbf{\Phi}_{c}^{*}\mathfrak{D}\mathbf{\Phi}_{c}
\mathfrak{D}\mathbf{\Omega}_{c} \int\mathfrak{D}\mathbf{\Phi_{i}}^{\prime\; * }
\mathfrak{D}\mathbf{\Phi}_{i}^{\prime}\mathfrak{D}\mathbf{\Omega}_{i}^{\prime} \;
\langle\mathbf{\Phi}_{i}^{\prime},\mathbf{\Omega}_{i}^{\prime} \left|  \;
\rho\; \right|  \mathbf{\Phi}_{i},\mathbf{\Omega}_{i} \rangle
\]
\[
\times\langle\mathbf{\Phi}_{i}, \mathbf{\Omega}_{i} \;\left|
{ \mathcal{T}^{\;-1}\exp}\left[  { -i\int d\mathbf{x\;}}\left(
\;{ \eta_{2}^{\ast}\left(  \mathbf{x}\right)  \cdot\mathbf{\Psi}\left(
\mathbf{x}\right)  +\mathbf{\Psi}^{+}\left(  \mathbf{x}\right)  \cdot\eta
_{2}\left(  \mathbf{x}\right)  +\mathbf{J}_{2}\left(  \mathbf{x}\right)
\cdot\mathbf{S}\left(  \mathbf{x}\right)  \;}\right)  \right]  \right|  \;
\mathbf{\Phi}_{c}, \mathbf{\Omega}_{c} \rangle
\]%
\begin{equation}
\; \;\times\langle\mathbf{\Phi}_{c}, \mathbf{\Omega}_{c} \left|
\mathcal{T}\exp\left[  \;i\int d\mathbf{x\;}\left(  \;{ \eta}_{1}^{\ast
}\left(  \mathbf{x}\right)  \cdot \mathbf{ \Psi}\left(  \mathbf{x}\right)
+\mathbf{ \Psi}^{+}\left(  \mathbf{x}\right)  \cdot{ \eta}_{1}\left(
\mathbf{x}\right)  +\mathbf{ J}_{1}\left(  \mathbf{x}\right)  \cdot
\mathbf{ S}\left(  \mathbf{x}\right)  \;\right)  \;\right]  \right|
\mathbf{\Phi}_{i}^{\prime}, \mathbf{\Omega}_{i}^{\prime}\;\rangle.
\end{equation}

\bigskip

\bigskip\noindent Hence we can now formally write the functional $\mathbb{Z}$
as a time-ordered path-integral around a closed path in time starting at $t =
t_{0}$, passing through $t = t_{c}$ and then going back to 
$t = t_{0}$ (see figure 2). \ This
functional then has a path-integral representation similar to the equilibrium
case \cite{Chou,Itzykson},
\begin{align}
\mathbb{Z}\left[  \eta^{\ast}, \; \eta, \; \mathbf{J}, \; \rho\right]   &  =
\int\mathfrak{D} \mathbf{\Psi}^{*} \mathfrak{D} \mathbf{\Psi} 
\mathfrak{D} \mathbf{m} \;
 \exp \left\{  \; i \; \mathcal{A} \left[  \mathbf{\Psi}^{*}, \; \mathbf{\Psi},
\; \mathbf{m},\; \eta^{*}, \; \eta, \; \mathbf{J} \right]  \; \right\}
\label{Zfun}\\
&  \; \; \; \; \; \; \times\langle\mathbf{\Psi}_{2},\mathbf{m}_{2}\left|  \;
\rho\; \right|  \mathbf{\Psi}_{1},\mathbf{m}_{1} \rangle\nonumber
\end{align}
where we have used the following notation for the now classical tensor fields
$\mathbf{m}$ and $\mathbf{\Psi}$,

\begin{align}
\mathbf{m}  &  \equiv\left(  \mathbf{m}_{+}, \; \mathbf{m}_{-} \right) \\
\mathbf{\Psi}  &  \equiv\left(  \mathbf{\Phi}_{+}, \; \mathbf{\Phi}_{-}
\right)
\end{align}
where $+$ and $-$ stand for the component that is propagating forward and
backward in time, respectively. The field $\mathbf{m}$ is therefore a
$2\times3$ tensor, while $\mathbf{\Psi}$ is a $2\times2 $ tensor. \ Similarly,
we write the source terms in terms of tensors. \ Using the following
notation for the external sources,
\begin{align}
\eta &  = \left(  \eta_{1}, \; \eta_{2} \right)  ,\\
\mathbf{J}  &  = \left(  \mathbf{J}_{1}, \; \mathbf{J}_{2} \right)  \;
.\nonumber
\end{align}
$\eta$ becomes a $2\times2$ tensor and $\mathbf{J}$ a $2\times3$ tensor.
\ This notation greatly simplifies the manipulation of the path-integral.

\begin{figure}
[ptb]
\begin{center}
\includegraphics[
natheight=8.770900in,
natwidth=6.198100in,
height=2.0859in,
width=4.4255in
]
{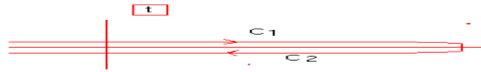}
\caption{Closed time path: branch 1 corresponds to forward 
propagation in time while branch 2 is that for backward propagation. }
\label{path}
\end{center}
\end{figure}

The action $\mathcal{A}$ is given as the difference of two actions; one due 
to
the fields propagating forward in time and the other due to fields propagating
backward in time,
\begin{align}
\mathcal{A}\left[  \mathbf{\Psi}^{*},\mathbf{\Psi} ,\mathbf{m}, \eta^{\ast
},\eta,\mathbf{J}\right]  = \mathcal{A}\left[  \mathbf{\Psi}^{*}%
_{1},\mathbf{\Psi}_{1} ,\mathbf{m}_{1}, \eta^{\ast}_{1},\eta_{1}
,\mathbf{J}_{1}\right]  - \mathcal{A}\left[  \mathbf{\Psi}^{*}_{2}%
,\mathbf{\Psi}_{2} ,\mathbf{m}_{2}, \eta^{\ast}_{2},\eta_{2} ,\mathbf{J}%
_{2}\right]  .
\end{align}
Both terms on the right are obtained in the usual way. \ The electron
contribution is standard. \ The magnetic moment contribution can be obtained
in the same way, but it involves a geometrical part coming from the $SU\left(
2 \right)  $ symmetry. \ Hence the forward part of the action is given by%

\begin{align}
\mathcal{A}\left[  \mathbf{\Psi}^{*}_{1},\mathbf{\Psi}_{1} ,\mathbf{m}_{1},
\eta^{\ast}_{1},\eta_{1} ,\mathbf{J}_{1}\right]  = \mathcal{A}_{WZ}\left[  \;
{\mathbf{m}}_{1}\; \right]  - \int d{\mathbf{x}} \; H_{d}\left(  {\mathbf{m}%
}_{1} \left(  \; \mathbf{x} \right)  \; \right) \\
\;\;\; + \int d\mathbf{x}\left\{  i\; \mathbf{\Psi} _{1 \;\alpha}^{\dagger
}\left(  \mathbf{x}\right)  \frac{\partial}{\partial t}\mathbf{\Psi} _{1
\;\alpha}\left(  \mathbf{x}\right)  - H_{s+sd}\left(  \mathbf{\Psi} _{1\;
\alpha}^{\dagger}, \; \mathbf{\Psi} _{1 \; \alpha}, \; \mathbf{m}_{1} \right)
\right\} \nonumber\\
\;\;\;\; +\int d \mathbf{x} \left\{  \eta_{1 \; \alpha}^{\ast} \left(
\mathbf{x} \right)  \mathbf{\Psi} _{1\; \alpha}\left(  \mathbf{x}\right)  +
\mathbf{\Psi} _{1 \; \alpha}^{\dagger}\left(  \mathbf{x}\right)  \eta_{1 \;
\alpha}\left(  \mathbf{x}\right)  +\mathbf{J}_{1}\left(  \mathbf{x}\right)
\cdot{\mathbf{m}}_{1}\left(  \mathbf{x}\right)  \right\}  ,\nonumber
\end{align}
where a summation over $\alpha$, the spin index, is implied. \ The
$\mathcal{A}_{WZ}$ is the Wess-Zumino part 
of the action $\mathcal{A}$. \ Because of the boundary conditions
on the spin fields at the left ends of the time path at $t=-\infty$ (KMS-type 
conditions), this WZ-term has the same form as in the equilibrium
case where the path of integration is along the 
imaginary-time branch from $t=0$ to $t=-i\beta$,\cite{fradkin,Rebei}
\begin{equation}
\mathcal{A}_{WZ}=\int_{0}^{1}d\tau \int_{C} dt \; \mathbf{m}(t,\tau)
\cdot \left[ \partial_{t}\mathbf{m}(t,\tau) \times 
\partial_{\tau}\mathbf{m}(t,\tau)
\right].
\end{equation}
The vector map $\mathbf{m}(t,\tau)$ is a parametrization of the 
surface enclosed by the trajectory of the magnetization, Fig. \ref{path}
\begin{eqnarray}
\mathbf{m}(t,0) &=&\mathbf{m}_{1}(t), t \in C_{1}  \\ 
&=&\mathbf{m}_{2}(t), t \in C_{2} , \nonumber \\
\mathbf{m}(t,1) &=& \mathbf{m}_{0},   \nonumber \\ 
\mathbf{m}(-\infty + i0^{+},\tau)& = &\mathbf{m}(-\infty + 
i0^{-},\tau). \nonumber \\
\end{eqnarray}
$\mathbf{m}_{0}$ is a distinguished vector and is usually taken 
along the quantization axis.

\ This WZ-term is topological in origin and 
can be considered as a constraint condition on the 
configuration space of
the magnetic moments and gives rise to magnetic monopole 
type potentials.
\cite{Bazaliy}

\bigskip

\noindent Next we expand the initial density matrix elements in terms of the
initial configurations of the conduction electron
 field and the magnetization field. \ Therefore we
are led to define a new functional $\mathbb{F}$ which may describe any initial
correlations between the conduction electrons  and the local magnetic 
moments,

\begin{equation}
\langle\mathbf{\Psi}_{2},\mathbf{m}_{2} \left|  \; \rho\; \right|
\mathbf{\Psi}_{1},\mathbf{m}_{1}\rangle=\exp\left\{  \; i\; \mathbb{F}\left[
\mathbf{\Psi}^{\dagger} , \mathbf{\Psi}, \; \mathbf{m} \right]  \; \right\}.
\end{equation}
Since we are assuming that the density matrix of the whole system is decoupled
at $t = t_{0}$, then all cross terms in the expansion are zero. \ Keeping 
only
terms up to second order, the expansion is

\begin{align}
\mathbb{F}\left[  \mathbf{m}, \; \mathbf{\Psi}, \; \mathbf{\Psi}^{\dagger}
\right]   &  = \mathbf{C}^{(0)} + \int d{\mathbf{x}} \; 
\epsilon^{\alpha\beta}
\mathbf{C}_{\alpha}^{(1)}\left(  \mathbf{x} \right)  \cdot\mathbf{m}_{\beta
}\left(  \mathbf{x} \right) \\
&  + \frac{1}{2} \int d{\mathbf{x}}d{\mathbf{y}} \; \epsilon^{\alpha\gamma}
\epsilon^{\beta\lambda} \mathbf{m}_{\alpha}\left(  \mathbf{x} \right)
\cdot\mathbf{C}^{(2)}_{\gamma\lambda} \left(  \mathbf{x}, \mathbf{y} \right)
\cdot\mathbf{m}_{\beta} \left(  \mathbf{y} \right) \nonumber\\
&  + \int d{\mathbf{x}}d{\mathbf{y}} \; \epsilon^{\alpha\gamma} \epsilon
^{\beta\lambda} \mathbf{\Psi}_{\alpha}^{\dagger}\left(  \mathbf{x} \right)
\cdot\mathbf{Q}_{\gamma\lambda} \left(  \mathbf{x}, \mathbf{y} \right)
\cdot\mathbf{\Psi}_{\beta}\left(  \mathbf{y} \right)  \; .\nonumber
\end{align}
The tensor $\epsilon$ is defined such that $\epsilon_{11} = - \epsilon_{22} =
1$, and $\epsilon_{12}=\epsilon_{21} = 0 $. \ The functional coefficients
$\mathbf{C}^{(0)}$, $\mathbf{C}^{(1)}$, $\mathbf{C}^{(2)}$, and $\mathbf{Q}$
are as yet unknown. \ The notation used should be clear; for example the last
term involves summations over the path index and the spin index,

\begin{align}
\epsilon^{\alpha\gamma} \epsilon^{\beta\lambda} \mathbf{\Psi}_{\alpha
}^{\dagger} \cdot\mathbf{Q}_{\gamma\lambda} \cdot\mathbf{\Psi}_{\beta} =
\epsilon^{\alpha\gamma} \epsilon^{\beta\lambda} {\Psi}_{s \; \alpha
}^{\dagger} {Q}_{\gamma\lambda}^{s \; s^{\prime}} 
{\Psi}_{s^{\prime} ,
\beta}
\end{align}
where the upper indexes on $\mathbf{Q}$ are for spin up and spin down.

\bigskip

\noindent Inserting this expansion back in Eq. (\ref{Zfun}), we 
end up with the
following expression for the action $\mathcal{A}$,
\begin{align}
\mathcal{A}\left[  \mathbf{\Psi}^{*},\mathbf{\Psi} , \mathbf{m},\eta^{\dagger
}, \eta, \mathbf{J}, \mathbf{Q}, \mathbf{C} \right]  = \epsilon^{\alpha\beta}
\left\{  \mathcal{A}_{WZ}\left[  \; {\mathbf{m}}_{\beta}\; \right]  - \int
d{\mathbf{x}} \; H_{d}\left(  {\mathbf{m}}_{\beta} \left(  \; \mathbf{x}
\right)  \; \right)  \right\} \\
+ \epsilon^{\alpha\beta}\int d\mathbf{x}\; \left\{  i\; \mathbf{\Psi} _{\beta
}^{\dagger}\left(  \mathbf{x}\right)  \frac{\partial}{\partial t}\mathbf{\Psi}
_{\beta}\left(  \mathbf{x}\right)  - H_{s+sd}\left(  \mathbf{\Psi} _{ \beta
}^{\dagger}, \; \mathbf{\Psi} _{ \beta}, \; \mathbf{m}_{\beta} \right)
\right\} \nonumber\\
+ \epsilon^{\alpha\beta} \int d \mathbf{x} \; \left\{  \eta_{ \beta}^{\dagger}
\left(  \mathbf{x} \right)  \mathbf{\Psi} _{ \beta}\left(  \mathbf{x} \right)
+ \mathbf{\Psi} _{ \beta}^{\dagger}\left(  \mathbf{x}\right)  \eta_{ \beta
}\left(  \mathbf{x}\right)  + \mathbf{J}_{\beta}\left(  \mathbf{x}\right)
\cdot{\mathbf{m}}_{\beta}\left(  \mathbf{x}\right)  \right\} \nonumber\\
+ \frac{1}{2}\epsilon^{\alpha\gamma}\epsilon^{\beta\lambda} \int d{\mathbf{x}%
}d{\mathbf{y}} \; \mathbf{m}_{\alpha}\left(  \mathbf{x} \right)
\cdot\mathbf{C}_{\gamma\lambda} \left(  \mathbf{x}, \mathbf{y} \right)
\cdot\mathbf{m}_{\beta} \left(  \mathbf{y} \right) \nonumber\\
+ \epsilon^{\alpha\gamma} \epsilon^{\beta\lambda}\int d{\mathbf{x}
}d{\mathbf{y}} \; \mathbf{\Psi}_{\alpha}^{\dagger}\left(  \mathbf{x} \right)
\cdot\mathbf{Q}_{\gamma\lambda} \left(  \mathbf{x}, \mathbf{y} \right)
\cdot\mathbf{\Psi}_{\beta}\left(  \mathbf{y} \right)  \; ,\nonumber
\end{align}
where we have made an obvious redefinition of the coefficients. The functional
$\mathbb{Z}$, now becomes of the standard form \cite{Itzykson}

\begin{align}
\mathbb{Z} \left[  \eta^{\dagger}, \eta, \mathbf{J}, \mathbf{Q}, \mathbf{C}
\right]  \;  &  = \;\\
&  \oint\mathfrak{D} \mathbf{\Psi}^{*} \mathfrak{D} \mathbf{\Psi} \mathfrak{D} \mathbf{m}
\; \exp\left\{  \; i \; \mathcal{A} \left[  \mathbf{\Psi}^{\dagger}, \;
\mathbf{\Psi}, \; \mathbf{m},\; \eta^{\dagger}, \; \eta, \; \mathbf{J}, \;
\mathbf{Q}, \; \mathbf{C} \right]  \; \right\}  . \nonumber\label{Zfun2}
\end{align}
The integral notation emphasizes that the path in time is closed, 
Fig. \ref{path}.
\ Therefore we now can apply the usual field theoretical methods to extract
the equations of motion from this functional.
\bigskip
From the correlation functions, it is clear that the functional
\begin{equation}
\mathbb{W}\;\left[  \;\eta^{\dagger},\;\eta,\;\mathbf{J}, \; \mathbf{Q}, \;
\mathbf{C}\;\right]  =\;-i\;\ln\mathbb{Z} \;\left[  \;\eta^{\dagger}%
,\;\eta,\;\mathbf{J}, \; \mathbf{Q}, \; \mathbf{C} \;\right]
\end{equation}
is the generator that we need to derive the irreducible Green's functions of
the system. \ To get the average value of the conduction 
electron field or the magnetization,
we differentiate with respect to the coefficients in the linear terms. \ For
the conduction electrons, we have
\begin{align}
\frac{\delta\mathbb{W} }{\delta\eta_{\alpha\; s}^{\dagger} \left(  \mathbf{x}
\right)  } = \; \epsilon^{\alpha\beta} \langle {\Psi}_{\beta\; s}
\left(  \mathbf{x} \right)  \rangle, \label{psi}
\end{align}
and for the magnetization,   we get
\begin{align}
\frac{\delta\mathbb{W} }{\delta {J}_{\alpha\; i} \left(  \mathbf{x}
\right)  } = \; \epsilon^{\alpha\beta} \langle {m}_{\beta\; i} \left(
\mathbf{x} \right)  \rangle. \label{m}
\end{align}
\bigskip The average of the conduction electrons
 field, a Fermi-type field,  
is set to zero while we set the average of the magnetization to be
\begin{equation}
{ {M}}_{\alpha\; i}\left(  \mathbf{x}\right)  =\langle m_{\alpha\;
i}\left(  \mathbf{x}\right)  \rangle.
\end{equation}
Given the above definitions, Eqs. (\ref{psi},\ref{m}), the two-point correlation
terms are easily obtained,

\begin{equation}
\frac{1}{\mathbb{Z}}\frac{\delta\mathbb{Z}}{\delta {Q}_{ \lambda\gamma
}^{s \; s^{\prime}}\left(  \mathbf{x}, \mathbf{y} \right)  } = i \; \frac
{\delta\ln\mathbb{W}}{\delta {Q}_{ \lambda\gamma}^{s \; s^{\prime}}\left(
\mathbf{x}, \mathbf{y} \right)  } = \; i \; \epsilon^{\alpha\lambda}
\epsilon^{\beta\gamma} \langle\Psi_{\alpha\; s }^{\dagger} \left(
\mathbf{x}\right)  \Psi_{\beta\; s^{\prime}}\left(  \mathbf{y} \right)  \rangle,
\end{equation}
\begin{align}
\frac{\delta\mathbb{W}}{\delta {C}_{\gamma\lambda}^{i j}\left(
\mathbf{x}, \mathbf{y} \right)  }\; = \; \epsilon^{\alpha\gamma}
\epsilon^{\beta\lambda} \;\frac{1}{2} \langle {m}_{\alpha\; i}\left(
\mathbf{x}\right)  {m}_{\beta\; j} \left(  \mathbf{\mathbf{y}} \right)
\rangle ,
\end{align}

\begin{align}
\frac{\delta\mathbb{W}}{\delta\eta_{\alpha\; s}\left(  \mathbf{x}\right)
\delta\eta_{\beta\; s^{\prime}}^{\dagger}\left(  \mathbf{y}\right)  } \; = \;
\epsilon^{\alpha\alpha^{\prime}}\epsilon^{\beta\beta^{\prime}}\langle
\Psi_{\alpha^{\prime}\; s }^{\dagger}(\mathbf{x})\Psi_{ \beta^{\prime} \;
s^{\prime}}(\mathbf{y})\rangle ,
\end{align}
where  $s$, $s^{\prime}$ are for spin
 up and spin down and $i$, $j$ are for the
spin field components. \ The indices $\alpha$, $\beta$ ..., denote 
the branch of time in fig. 2.
\ Mixed correlation functions can be obtained in the same way.

\bigskip

Clearly, solving for the two-point propagators is the 
least we can do to have
a meaningful solution that includes relaxation effects. \ Knowing 
these
propagators amounts to knowing the particle density, the spin 
density, the
current density, and the scattering amplitudes, among others. 
 \ Since we assume that the conduction 
electrons' field has no mean
value, its two-point propagator is then explicitly given by time-ordered 
products,

\begin{align}
\mathcal{G}^{ss^{\prime}}_{22}\left(  \mathbf{x},\mathbf{y}\right)   &
=\langle\; \mathcal{T}^{-1}\left(  \Psi_{s}\left(  \mathbf{x}\right)
\Psi_{s^{\prime}}^{+}\left(  \mathbf{y}\right) \; \right)  \rangle ,\\
\mathcal{G}^{ss^{\prime}}_{21}\left(  \mathbf{x},\mathbf{y}\right)   &
=\langle  \Psi_{s}\left(
\mathbf{x}\right)  \Psi_{s^{\prime}}^{+}\left(  \mathbf{y}\right)  
 \rangle , \nonumber\\
\mathcal{G}^{ss^{\prime}}_{11}\left(  \mathbf{x},\mathbf{y}\right)   &
=\langle\; \mathcal{T}\left(  \Psi_{s}\left(  \mathbf{x}\right)
\Psi_{s^{\prime}}^{+}\left(  \mathbf{y}\right)  \right)  \; 
\rangle ,  \nonumber\\
\mathcal{G}^{ss^{\prime}}_{12}\left(  \mathbf{x},\mathbf{y}\right)   &
= - \langle  \Psi_{s^{\prime}}^{+}\left(
\mathbf{y}\right)     \Psi_{s}\left(  \mathbf{x}\right)   
 \rangle. \nonumber
\end{align}
From the above expressions, it is clear that the 
function $\mathcal{G}_{21}$
is the ``less than'' Green's function and $\mathcal{G}_{12}$ is the ``greater
than'' Green's function. $\mathcal{G}_{11}$ is the Feynman propagator, while
$\mathcal{G}_{22}$ is the Dyson propagator.
\cite{Ferry} \ These Green's functions are not all independent. \ From 
their definitions, we can see that 
\begin{equation}
\mathcal{G}^{ss^{\prime}}_{11}\left(  \mathbf{x},\mathbf{y}\right) +
\mathcal{G}^{ss^{\prime}}_{22}\left(  \mathbf{x},\mathbf{y}\right) =
\mathcal{G}^{ss^{\prime}}_{12}\left(  \mathbf{x},\mathbf{y}\right)  +
\mathcal{G}^{ss^{\prime}}_{21}\left(  \mathbf{x},\mathbf{y}\right). 
\end{equation}
The Green's function $\mathcal{G}_{12}$ is of special interest since 
it is related to the average of the density operator of the conduction
electrons.  \ The two-point functions for the magnetization are similarly
given by

\begin{align}
\mathcal{M}^{ij}_{22}\left(  \mathbf{x},\mathbf{y}\right)   &  =\langle\;
\mathcal{T}^{-1}\left(  {S}_{i}\left(  \mathbf{x}\right)  \right)  \left(
S_{j} \left(  \mathbf{y} \right)  \right)  \rangle - \langle S_{i} \left(
  \mathbf{x}\right) \rangle \langle S_{j} \left(
  \mathbf{y}\right) \rangle ,   \\
\mathcal{M}^{ij}_{21}\left(  \mathbf{x},\mathbf{y}\right)   &  =\langle
S_{j}\left(  \mathbf{y}\right)  S_{i}\left(  \mathbf{x} \right)
\rangle - \langle S_{i} \left(
  \mathbf{x}\right) \rangle \langle S_{j} \left(
  \mathbf{y}\right) \rangle  ,    \nonumber\\
\mathcal{M}^{ij}_{11}\left(  \mathbf{x},\mathbf{y}\right)   &  =\langle
\mathcal{T}\left(  S_{i}\left(  \mathbf{x}\right)  S_{j}\left(  \mathbf{y}
\right)  \right)  \rangle - \langle S_{i} \left(
  \mathbf{x}\right) \rangle \langle S_{j} \left(
  \mathbf{y}\right) \rangle , \nonumber\\
\mathcal{M}^{ij}_{12}\left(  \mathbf{x},\mathbf{y}\right)   &  =\langle
S_{i}\left( \mathbf{x}\right)  S_{j}\left(  \mathbf{y}\right)  \rangle
 - \langle S_{i} \left(
  \mathbf{x}\right) \rangle \langle S_{j} \left(
  \mathbf{y}\right) \rangle . \nonumber
\end{align}
\ These Green's functions are easily related to the retarded and advanced
Green's functions. \ Since we are considering a situation which is not far
from equilibrium, we will follow closely the treatment in Ref. 
\onlinecite{Baym}.
\ Therefore, as in the equilibrium case, we relate the ``less than'' functions
to the distribution function of electrons and spin in the semi-classical 
limit.

\bigskip

\bigskip

\section{The Variational Principle: Effective action method}

\bigskip

\begin{figure}
[ptb]
\begin{center}
\includegraphics[
natheight=8.770900in,
natwidth=6.198100in,
height=2.0859in,
width=4.4255in
]
{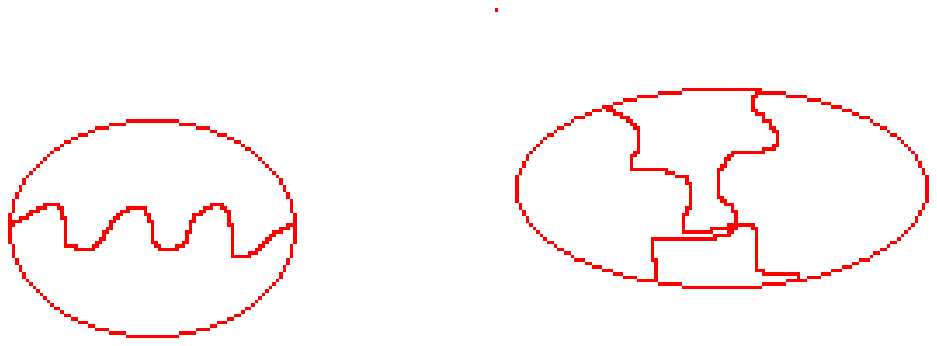}
\caption{Interaction terms between the conduction electrons  and the localized 
magnetic moments. The
smooth curve represents the conduction electron propagator. The curved line 
represents the
spin-spin correlation functions.}
\label{loops}
\end{center}
\end{figure}

Since the functions $\mathbf{J}$, $\mathbf{Q}$, $\mathbf{C}$ are not bound 
to
a simple physical interpretation, we make the following Legendre 
transformation,

\bigskip
\begin{align}
\Gamma\left[  \mathbf{M}_{\alpha\; i }\left(  \mathbf{x} \right)  ,
{\mathcal{G}}_{\alpha\beta}^{s s^{\prime}}\left(  \mathbf{x},
\mathbf{y}\right)
,{\mathcal{M}}_{\alpha\beta}^{i j}\left(  \mathbf{x},\mathbf{y}\right)
\right]   &  = \mathbb{W}\left[  \mathbf{J}_{\alpha\; i}, \mathbf{Q}
_{\alpha\beta}^{s s^{\prime}}, \mathbf{C}_{\alpha\beta}^{i j}\right]  -\int
d\mathbf{x} \; {J}_{\alpha\; i} \left(  \mathbf{x}\right)
 {M}_{\alpha\; i}\left(  \mathbf{x}\right) \\
&  -\int d\mathbf{x}d\mathbf{y} \; {Q}_{\alpha\beta}^{s s^{\prime}}
\left(  \mathbf{x},\mathbf{y} \right)  {\mathcal{G}}_{\beta\alpha}^{s^{\prime}
s}\left(  \mathbf{y,x}\right) \nonumber\\
&  -\frac{1}{2}\int d\mathbf{x}d\mathbf{y}\; {C}_{\alpha\beta}^{ij}
\left(  \mathbf{x},\mathbf{y}\right)  \left(  \mathcal{M}_{\beta\alpha}^{j
i}\left(  \mathbf{y,x}\right)  + {M}_{\beta\; j}\left(  \mathbf{y}%
\right)  {M}_{\alpha\; i }\left(  \mathbf{x}\right)  \right)
.\nonumber
\end{align}
We end up with a functional $\Gamma$ that is expressed solely in terms of
magnetization and correlation functions of the current and the localized
spins. \ The equations of motion are then found by differentiating $\Gamma$
with respect to its arguments,

\begin{align}
\frac{\delta\Gamma}{\delta {M}_{\alpha\; i}\left(  \mathbf{x} \right)
}  &  = - {J}_{\alpha\; i} \left(  \mathbf{x} \right)  - \int
d{\mathbf{y}} \; {C}_{\alpha\beta}^{ij} \left(  \mathbf{x},
\mathbf{y}\right)  {M}_{\beta\; j} \left(  \mathbf{y} \right)  ,\\
\frac{\delta\Gamma}{\delta\mathcal{G}_{\alpha\beta}^{ss^{\prime}}\left(
\mathbf{x}, \mathbf{y} \right)  }  &  = - {Q}_{\beta\alpha}^{s^{\prime
}s}\left(  \mathbf{y}, \mathbf{x} \right)  ,\\
\frac{\delta\Gamma}{\delta\mathcal{M}_{\alpha\beta}^{ij}\left(  \mathbf{x},
\mathbf{y} \right)  }  &  = - \frac{1}{2} {C}_{\beta\alpha}^{ji}\left(
\mathbf{y}, \mathbf{x} \right)  .
\end{align}
Using the standard tools of field theory \cite{Jackiw}, we solve for
$\mathbf{J}$, $\mathbf{Q}$, $\mathbf{C}$ in terms of $\mathbf{M}$,
$\mathcal{G}$ and $\mathcal{M}$. \ A discussion of Wick's theorem 
is beyond the scope of this paper.
\ Omitting terms of 
 order $\lambda^{4}$ and higher,  we have
the approximate effective action for the conduction electrons 
and the localized magnetic moments,

\begin{align}
\Gamma\left[  {\mathbf{M}},{\mathcal{G}},{\mathcal{M}}\right]   &
=\mathcal{A}_{0}\left[  {\mathbf{M}}\right]  +\frac{i}{2}
\ln\det{\mathcal{M}
}^{-1}\\
&  +\frac{1}{2}\int d\mathbf{x}d\mathbf{y}\;
\left[ \frac{\delta^{2}\mathcal{A}_{0}
}{\delta {M}_{\alpha\;i}\left(  \mathbf{x}\right) 
 {M}_{\beta
\;j}\left(  \mathbf{y}\right)  }{\mathcal{M}}_{\beta\alpha}^{ji}\left(
\mathbf{y},\mathbf{x}\right)\right]  -\;i\;\ln\det{\mathcal{G}}^{-1}
\nonumber\\
&  +\int d\mathbf{x}d\mathbf{y}\;
\left[ \frac{\delta^{2}\mathcal{A}_{0}}
{\delta {\Psi}_{\alpha\;s}^{\dagger}\left(  \mathbf{x} \right)
\delta {\Psi}_{\beta\;s^{\prime}}\left( \mathbf{y}\right)}{\mathcal{G}
}_{\beta\alpha}^{ss^{\prime}}\left(  \mathbf{y},
\mathbf{x}\right) \right]  \nonumber\\
&  +\frac{\lambda^{2}}{2}g^{\alpha\alpha^{\prime}\alpha^{\prime\prime}
}g^{\beta\beta^{\prime}\beta^{\prime\prime}} \frac{\sigma_{s_{4}s_{1}}^{i}}{2}
\frac{\sigma_{s_{2}s_{3}}^{j}}{2}\int d\mathbf{x}d\mathbf{y}\;
\left[ {\mathcal{G}}_{\alpha
\beta}^{s_{1}s_{2}}\left(  \mathbf{x},\mathbf{y}\right)  
{\mathcal{M}}
_{\alpha^{\prime\prime}\beta^{\prime\prime}}^{ij}\left(  \mathbf{y}
,\mathbf{x}\right)  {\mathcal{G}}_{\alpha^{\prime}\beta^{\prime}}^{s_{3}s_{4}
}\left(  \mathbf{y},\mathbf{x}\right) \right]  \nonumber\\
&  + O( \lambda^{4}) .\nonumber
\end{align}
\bigskip
\noindent The functional $\mathcal{A}_{0}$ is the functional $\mathcal{A}$ 
with all
the source terms set to zero. \ The tensor $g^{ijk}$ is equal to 1 if
$i=j=k=1$ and equal to -1 if $i=j=k=2$ and zero, otherwise. \ The last 
term, which is clearly valid 
for large magnitude of $\mathbf{S}$,  has a simple interpretation in terms of 
Feynman diagrams, Fig. 
\ref{loops}.

The equations of motion for $\mathbf{M}$, $\mathcal{G}$, $\mathcal{M}$, are
obtained by minimizing $\Gamma$ and setting the external sources to zero with
the appropriate initial conditions.  \ Within the above stated 
approximations,  the magnetization of 
the medium obeys the following equation of motion
\begin{equation}
\begin{array}
[c]{cc}
\epsilon^{\alpha\beta}\;\epsilon^{ilk}\;{M}_{\beta l}\left(
\mathbf{x}\right)  
\partial_{t} {M}_{\beta k}\left(  \mathbf{x}\right)
+\epsilon^{\alpha\beta}\;\delta_{\alpha\beta}\;B_{i}\left(
\mathbf{x}\right)   & \\
+\frac{1}{2}\;J\;\epsilon^{\alpha\beta}\;
\nabla^{2} {M}_{\beta
i}\left(  \mathbf{x}\right)  + \frac{\lambda}{2}\;\epsilon^{\alpha^{\prime}
\alpha}
\sigma_{s^{\prime}s}^{i}\mathcal{G}_{\alpha^{\prime}\alpha}^{ss^{\prime}
}\left(  \mathbf{x},\mathbf{x^{\alpha}}\right)   & =0 .
\end{array}\label{eq57}
\end{equation}
Here we have taken the long-wavelength limit to get the familiar 
exchange term
through a coarse-graining procedure where each cell is taken to 
have a maximum
spin of $S$. \ The last term on the left is clearly the interaction 
with the
conduction electrons'  magnetic moments to {\it{all}} orders in $\lambda$. \ The 
equation of motion
for the conduction electrons is

\begin{eqnarray}
\left[ \left(
i\partial_{{t}_{y}}-\in (y)\right)\delta_{s^{\prime}s}
  + \frac{\mu}{2}{
\sigma_{s^{\prime}s}^{i}} {B}^{i} + \frac{\lambda}{2} \sigma_{ss^{\prime}}
  {M}_{\alpha i}\left( \mathbf{y} \right)    \right]  
\mathcal{G}_{\gamma\alpha
}^{s s^{\prime\prime}}\left( \mathbf{y}, \mathbf{z}\right) 
\;\;\;\;\;\;\;\;\;\;  &&  \label{eq2}\\
+\lambda^{2}g^{\alpha\alpha^{\prime}\alpha^{\prime\prime}}g^{\beta
\beta^{\prime}\beta^{\prime\prime}}\frac{\sigma_{s_{4}s}^{i}}{2}
\frac{\sigma_{s^{\prime}s_{3}
}^{j}}{2}\int d\mathbf{x}
\left[ \mathcal{G}_{\alpha^{\prime}\beta^{\prime}}^{s_{3}s_{4}
}\left(  \mathbf{y},\mathbf{x}\right)  \mathcal{G}_{\gamma\alpha}
^{s^{\prime\prime}s}\left(  \mathbf{z},\mathbf{x}\right)  \mathcal{M}
_{\alpha^{\prime\prime}\beta^{\prime\prime}}^{ij}\left(  \mathbf{x}
,\mathbf{y}\right) \right]  & & = - i\delta_{\gamma\beta}^{s^{\prime}
s^{\prime\prime}}\left( \mathbf{z}-\mathbf{y}\right) \;,\nonumber
\end{eqnarray}
where $\in (y)$ is the spin-independent energy of the conduction
electron.  \  The term of first order in $\lambda$  describes  
 the {\it{full}}
exchange interaction between the magnetic moments of the localized 
electrons
and those of the current. \ The structure of this equation is 
familiar from the theory of correlation functions due to Coulomb 
interactions.\cite{rebei3} \ There, the propagator $\mathcal{M}(
\mathbf{x},\mathbf{y})$ is replaced by the Hartree propagator. \ Therefore 
the solution of this equation should follow by analogy with 
the treatment in Ref. \onlinecite{rebei3}.
\ The final equation is the equation of 
motion for
the magnetic correlation functions,
\begin{eqnarray}
 \epsilon^{\alpha\beta}\epsilon^{ijk}\left\{  
\partial_{t} {M}_{\beta
k}\left(  \mathbf{x}\right)  - {M}_{\beta k}\left(  
\mathbf{x}\right)
\partial_{t}\right\}  {\mathcal{M}}_{\alpha^{\prime}\beta}^{k^{\prime}
j}\left(  \mathbf{z},\mathbf{x}\right)    
+ \epsilon^{\alpha\beta}\int d\mathbf{y}\;
\left[ J\left(  \mathbf{x}
-\mathbf{y}\right) 
 {\mathcal{M}}_{\alpha^{\prime}\beta}^{k^{\prime}i}\left(
\mathbf{z},\mathbf{y}\right) \right]  &&   \label{eq3}\\
+\lambda^{2}g^{\alpha_{1}\alpha_{2}\beta}g^{\beta_{1}\beta_{2}\alpha}
\frac{\sigma_{s_{4}s_{1}}^{j}}{2}\frac{\sigma_{s_{2}s_{3}}^{i}}{2}
\int d\mathbf{y}
\;\left[ {\mathcal{M}
}_{\alpha^{\prime}\beta}^{k^{\prime}j}\left(  \mathbf{z},\mathbf{x}\right)
\mathcal{G}_{\alpha_{1}\beta_{1}}^{ss^{\prime}}\left(  \mathbf{x},\mathbf{y}
\right)  \mathcal{G}_{\alpha_{2}\beta_{2}}^{s^{\prime}s}\left(  \mathbf{y,x}
\right)  \right] &  = & i \delta_{\alpha^{\prime}\alpha}^{k^{\prime}i}\left(
\mathbf{x}-\mathbf{z}\right)  \;.\nonumber
\end{eqnarray}
The integrals are all 
four-dimensional and hence we have defined $J \left(\mathbf{x} - 
\mathbf{y} \right) = J \left(\overrightarrow{\mathbf{x}} - 
\overrightarrow{\mathbf{y}} \right)\delta \left( t_{x} - t_{y} 
\right)$. \  
This set of coupled equations, Eqs. $( \ref{eq57}-\ref{eq3} )$,
 is the main result of this work. \ It 
is
important to observe that up to   this point, 
 the propagators $\mathcal{M}$ 
and
$\mathcal{G}$ are the \textit{{true}} propagators of the theory. \ Hence the
above equations are non-perturbative in nature. \ In this 
section and the  next, we 
study the structure of  
 these equations  and make contact with 
previous work. \ Since
we are interested in how the magnetic moment of the current 
influences that of
the medium (or vice-versa), we define

\begin{eqnarray}
\mathfrak{M}_{\alpha\beta}^{i}\left(  \mathbf{x},\mathbf{y}\right)  
\; & = & \;\frac
{1}{2}\sigma_{ss^{\prime}}^{i}\mathcal{G}_{\alpha\beta}^{s^{\prime}s}\left(
\mathbf{x},\mathbf{y}\right)  \;,
\end{eqnarray}
to be the conduction electron spin 
 propagator. \ The spin ``charge''  of the current is easily
seen to follow from $\mathfrak{M}_{\alpha\beta} $ by setting $\alpha=1$, $\beta=2$ and letting
$\mathbf{y}\rightarrow\mathbf{x}^{+}$,
\begin{equation}
\mathfrak{M}^{i}\left(  \mathbf{x}\right)  = \frac{1}{2}\sigma_{ss^{\prime}}
^{i}\mathcal{G}_{12}^{s^{\prime}s}\left(  \mathbf{x},\mathbf{x}\right)  .
\end{equation}
However, we will find it useful to go to the center of mass and relative 
coordinates (Wigner coordinate system) to 
make contact with the classical quantities,
\begin{eqnarray}
\mathbf{X} & = & \frac{1}{2} \left( \overrightarrow{\mathbf{x}} + 
\overrightarrow{\mathbf{y}} \right) , \\
T & = &  \frac{1}{2} \left( t_{x} + t_{y} \right) , \nonumber \\
\mathbf{x}_{\Delta} & = & \overrightarrow{\mathbf{x}} -
\overrightarrow{\mathbf{y}} ,  \nonumber \\
t_{\Delta} & = &  t_{x} - t_{y} \; .  \nonumber
\end{eqnarray}
In this new coordinate system, the magnetic 
moment of the conduction electrons  $\mathfrak{M}^{i}(\mathbf{x})$ 
 becomes a function of the macroscopic variables 
$\mathbf{X}$ and $T$ only,
\begin{equation}
\mathfrak{M}^{i}(\mathbf{x}, \mathbf{y} )=\int \frac{d\omega}{2 \pi} 
\int \frac{d \mathbf{p}}{(2 \pi)^{3}} \exp\left[i \omega t_{\Delta} 
- i \mathbf{p}\cdot \mathbf{x}_{\Delta}  \right]
\mathfrak{M}^{i}( \mathbf{X},T; \omega,
 \mathbf{p} ).
\end{equation}

To get the equation of motion for  $\mathfrak{M}$,
    the spin charge of the conduction electrons, we first
multiply Eq. (\ref{eq2}) from the L.H.S. by $\sigma^{l}_{ss^{\prime\prime}}$ 
and sum over the spin degrees of freedom. \ We end up with 
an equation for the polarized current propagator,
\begin{eqnarray}
\epsilon^{\alpha \beta}\;\delta_{ss^{\prime}}\left[
i\partial_{{t}_{y}}-\in_{\alpha} \right]  \mathbf{\mathfrak{M}}_{\gamma\alpha
}\left(  \mathbf{y},\mathbf{z}\right)
+ \frac{ i \epsilon^{\alpha \beta}}{2} \left[  \mathbf{B} + \lambda 
\mathbf{M}\left( \mathbf{y}
\right)   \right] \times  
 \mathbf{ \mathfrak{M}}_{\gamma\alpha
}\left(  \mathbf{y},\mathbf{z}\right) \label{current}   \\
+\lambda^{2}g^{\alpha\alpha^{\prime}\alpha^{\prime\prime}}g^{\beta
\beta^{\prime}\beta^{\prime\prime}}   
 \frac{\sigma_{s_{4}s}^{i}}{2} \frac{\sigma_{s^{\prime}s_{3}
}^{j}}{2} \int d\mathbf{x}\left[ 
\mathcal{G}_{\alpha^{\prime}\beta^{\prime}}^{s_{3}s_{4}
}\left(  \mathbf{y},\mathbf{x}\right)  
  \mathcal{M}
_{\alpha^{\prime\prime}\beta^{\prime\prime}}^{ij}\left(  \mathbf{x}
,\mathbf{y}\right) \sigma^{l}_{s^{\prime\prime}s^{\prime}}\mathcal{G}_{\gamma\alpha}^{s^{\prime\prime}s}\left(  \mathbf{z},\mathbf{x}\right)
\right]  &  = 0  \;.\nonumber
\end{eqnarray}
\bigskip
The last term on the left provides for the relaxation of 
the spin moment 
of the conduction electrons.

\ To derive the equation 
for the polarization of the 
current, in the following we use the relaxation time approximation 
and replace the last term in Eq. (\ref{current}) , the 
collision integral, by a local term. \ The classical
polarization of the current $\mathbf{\mathfrak{M}_{c}}$
is found by first assuming that the $l-$th component
, $\mathfrak{M}^{l}(\mathbf{x})$, has the following
form

\begin{equation}
\mathfrak{M}^{l}\left(  T,X,p,\omega\right)  =\delta\left(  \omega-p^{2}%
/2m-V\left(  T,X\right)  \right)  \mathfrak{M}^{l}\left(  T,X,p\right),
\end{equation}
where we have set $\in (p) = p^{2}/2m + V(T,X)$. \
Then by averaging over the fast degrees of freedom, we have by 
definition

\begin{equation}
\mathfrak{M}_{c}^{l}\left(  T,X\right)  = v \int \frac{d\omega}{2\pi}
\frac{d\mathbf{p}}{\left(
2\pi\right)^{3}} \; \mathfrak{M}^{l}\left(  T,X,p,\omega\right),
\end{equation}
where $v$ is the volume of the system.
The spin current $\mathbf{\mathfrak{J}}$ is defined 
in the usual way. \ However here 
it has a tensorial character because of the vector character 
of the spin charge,

\begin{equation}
\mathfrak{J}^{kl}\left(  T,X\right)  =\frac{v}{m}\int\frac{d\mathbf{p}}{(2\pi
)^{3}} \; \mathfrak{M}^{k}\left(  T,X,p\right)  p^{l}.
\end{equation}
The equation of motion for $\mathbf{\mathfrak{M}}_{c}$ is found by
first going to the center of mass coordinates and using 
the quasi-particle approximation.\cite{Ferry} \ We find for each 
$k-$component

\begin{eqnarray}
\left[  \partial_{T}+\frac{p}{m}\cdot\partial_{X}\right]  
\mathfrak{M}^{k}\left(
T,X,p\right) & +& \epsilon^{klp}\left[  B^{l}+\lambda M^{l}\left(  T,X\right)
\right]  \mathfrak{M}^{p}\left(  T,X,p\right) \nonumber \\
& =&-\frac{\mathfrak{M}^{k}\left(
T,X,p\right)  -\mathfrak{M}_{eq}^{k}  }{\tau_{k}}
-\frac{\mathfrak{M}^{k}\left(
T,X,p\right)  -\mathfrak{M}_{0}^{k}  }{\tau_{p}} , \label{current2}
\end{eqnarray}
where $\tau_{k}(\tau_{p})$, is the relaxation time for spin flip 
(momentum) scattering 
processes.\cite{davies,rammer}
\ By definition the average of the last term over the 
momentum is zero. \ This way of writing the collision term is 
valid only in the 
absence of spin-momentum coupling terms such as $\mathbf{L}\cdot\mathbf{S}$-coupling. 

\bigskip

\ It is worthwhile to pause here and consider the content of this 
equation. \ The first term on the left hand side
is the total time derivative, with independent variables 
$(T,\mathbf{X},\mathbf{p})$ and $d\mathbf{p}/dt=0$ ( we consider 
a non-zero electric field elsewhere \cite{rebei4} ). \ The 
other term on the left hand side is the torque due to the 
local moments and the right hand side is 
an approximate expression for the collision operator.

\bigskip

\ In the
absence of gradients, the polarization of the conduction electrons is along 
the local effective field and there will be no spin currents. \ This 
will not be the case if there is an electric field present. \ The  spin 
accumulation effects studied here are solely due to non-homogeneous
magnetization of the medium. \ It is important to observe that this effect
is present even in the absence of a current as is the case for the spin 
accumulation effects due 
to interfaces put forward by Berger.\cite{Berger} \ In fact 
paramagnetic-ferromagnetic interfaces are naturally included in our 
treatment since we are dealing with nonhomogeneous magnetization. 
 \ An interface simply corresponds to an abrupt change of the 
magnetization from a non-zero value to a zero value. \ These particular 
effects will be treated below. 

\bigskip

\ Next 
we multiply
 Eq. $( \ref{current2})$ by the velocity 
and then average over it. \ To obtain the usual  Fick's law for 
spin diffusion, we assume that the momentum
  relaxation time is  small and 
hence the R.H.S of Eq. (\ref{current2}) is larger than the effects due
to the local magnetization $\mathbf{M}$. \ In this case we have 

\begin{equation}
\mathfrak{J}^{kl}\left(  T,X\right)  =-D^{k}\partial_{X_{l}}\mathfrak{M}_{c}%
^{k}\left(  T,X\right),\label{ficks}
\end{equation}
where the diffusion constants are given in terms of an 
average Fermi velocity
$ v_{F}$ by
\begin{equation}
D^{k}=\frac{1}{3}v_{F}^{2}\tau_{k}^{eff},
\end{equation}
where $1/{\tau_{k}^{eff}}=1/{\tau_{k}} + 1/{\tau_{p}}$. \
To get this result we made use of the following approximation
for the velocities of the conduction electrons
\begin{equation}
\partial_{X_{l}}\int\frac{d\widehat{p}}{4\pi}v^{l}v^{j}\mathfrak{M}^{k}\left(
T,X,p\right)  \simeq\frac{1}{3}v_{F}^{2}\partial_{Xj}\mathfrak{M}_{c}^{k}\left(
T,X\right).
\end{equation}
Finally using Eqs.(\ref{current2}, \ref{ficks}), we find that the 
classical magnetization of the 
conduction electrons obeys a diffusion equation for each one of 
its components,

\begin{equation}
\left[  \partial_{T}-D^{k}\nabla^{2}\right]  \mathfrak{M}_{c}^{k}\left(
T,X\right)  =-\frac{1}{\tau_{k}}\left(  \mathfrak{M}_{c}^{k}\left(  T,X\right)
-\mathfrak{M}_{eq}^{k}\left(  X\right)  \right)  -\left[  \left(  \mathbf{B}+
\lambda \mathbf{M}\right)
\times\mathfrak{M}_{c}\right]  ^{k}.
\end{equation}
This equation is however rotationally invariant and does not show the 
reduced symmetry of the ferromagnetic state. \ To get a more realistic 
equation we improve on Fick's law by keeping all terms in Eq.(\ref{current2}) and treat exchange 
effects between the conduction electrons and the magnetization more carefully. \ This amounts to 
taking into account the sd-exchange term in the electron propagators. 
\ For slow variations in time, we have now a modified Fick's law 
that takes into account the variation of local magnetization in space 
and in direction,
\begin{equation}
\mathfrak{J}^{kj}\left(  T,\mathbf{X}\right)  =-\mathfrak{D}^{kp}\partial_{X_{j}%
}\mathfrak{M}_{c}^{p}\left(  T,\mathbf{X}\right), \label{flux}
\end{equation}
where now the diffusion constant becomes a tensor. \ It is defined in 
terms of a matrix $A$
\begin{equation}
\mathfrak{D}^{kp}\left(  \mathbf{X}\right)  =D^{p}\left(  A^{-1}\right)^{kp}(
\mathbf{X}). \label{tensor}
\end{equation}
There is no summation over $p$ in this equation.
The matrix $A$ depends locally on the effective magnetization 
field $\mathbf{H}$,
\begin{equation}
A\left(  \mathbf{X}\right)  =\left[
\begin{array}
[c]{ccc}%
1 & -\tau_{x}H_{z} & \tau_{x}H_{y}\\
\tau_{y}H_{z} & 1 & -\tau_{y}H_{x}\\
-\tau_{z}H_{y} & \tau_{z}H_{x} & 1
\end{array}
\right].
\end{equation}
In our approximation, the effective local field is simply 
\begin{equation}
\mathbf{H}=\mathbf{B}+\lambda\mathbf{M} \left( \mathbf{x} \right).
\end{equation}
Now in the steady state, the equation satisfied by the average magnetic
moment
$\mathfrak{M}_{c}$ 
becomes a generalized diffusion equation

\begin{equation}
\sum_{p,l}\partial_{X_{l}}\left[  \mathfrak{D}^{kp}\left(  X\right)
\partial_{X_{l}}\mathfrak{M}_{c}^{p}\left(  \mathbf{X}\right)  \right]  =\frac
{1}{\tau_{k}}\left(  \mathfrak{M}_{c}^{k}-\mathfrak{M}_{eq}^{k}\right)  +\lambda
\left[  \mathbf{M}\times\mathfrak{M}_{c}\right]  ^{k} \label{diffusion}
\end{equation}
The tensor character of the diffusion term in Eq. $(\ref{diffusion})$ is 
not due to anisotropic transport - the flux in the $j-$direction
in Eq. $(\ref{flux})$ is due to a gradient with respect to $X_{j}$. \ Rather, 
the p-component of $\mathfrak{M}_{c}$ is rotated into the $k-th$ direction 
by the effective field $\mathbf{H}$, while transport takes place in the 
direction of  the  gradient. \ The diffusion 
tensor, Eq.(\ref{tensor}), has striking similarities to the diffusion tensor 
of charged species in a   plasma.  
\cite{bittencourt} \ If we restrict ourselves to the case where the 
local effective field is constant and 
along the $z-$axis only, then the transverse 
diffusion coefficients are similar to those  found by Hirst 
\cite{hirst} and Kaplan \cite{kaplan} using 
very different methods from the one 
presented here. \ Their work showed that in the 
direction perpendicular to the effective field, diffusion 
of polarization of the electron gas is much slower than 
along the field. \ The off-diagonal 
terms have their origin in the strong exchange interaction 
among the conduction electrons which can not be treated 
perturbatively for a transition metal. \ In a magnetic metal such 
as Ni, the off-diagonal terms can be two orders of magnitude 
larger than the diagonal ones.\cite{hirst}

\bigskip

\section{Spin-Momentum Transfer: A self-consistent treatment}

\bigskip

\begin{figure}
[ptb]
\begin{center}
\includegraphics[
natheight=8.7027700in,
natwidth=6.127700in,
height=4.0277in,
width=4.0277in
]
{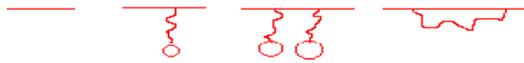}

\caption{Series expansion of the conduction electron
 propagator in powers of 
$\lambda$. }
\label{green}
\end{center}
\end{figure}

To solve the above equations of motion, we retain a subset of the 
terms arising in the full spin propagator. \ This approximation 
is essentially similar to the random phase approximation 
in the calculation of the ground state energy of an interacting 
electron gas.\cite{rebei3} \ The zero
order propagator is taken to be that of the electrons  in the 
external $\mathbf{B}$-field  and the localized spins interacting through the
exchange interaction in the presence of the $\mathbf{B}$ 
field. \ Since the magnetic moments constitute
 a many-body problem, a full solution is not possible in general. Hence
an explicit solution to the problem requires first a calculation of 
the background magnetization. \ The magnetization  satisfies a 
generalized Landau-Lifshitz equation which follows from Eq.
(\ref{eq57}). First we 
observe that when the external sources are turned off, we have
\begin{equation}
\mathbf{M}_{1}\left(  \mathbf{x}\right)  =\mathbf{M}_{2}\left(  \mathbf{x}
\right)  =\mathbf{M}\left(  \mathbf{x}\right)  ,
\end{equation}
where $\mathbf{M}$ is the average, i.e. classical,  magnetization. Equation (\ref{eq57}) 
is a system of
two equations for $\mathbf{M}_{1}$ and $\mathbf{M}_{2}$, the 
magnetization vector along the paths $C_{1}$ and $C_{2}$ 
respectively, Fig. \ref{path}. \ It is the 
averaging
of these two equations that gives the equation of motion for 
the average
magnetization,
\begin{equation}
\partial_{t}\mathbf{M}\left(  \mathbf{x}\right)  =\mathbf{M}\left(
\mathbf{x}\right)  \times\left[  \frac{1}{2} J \; \nabla^{2}\mathbf{M}
\left(
\mathbf{x}\right)  {+}  \mathbf{B} + \frac{\lambda}
{2}\overrightarrow{\mathbf{\sigma}}_{s^{\prime}s}\frac{1}{2}\left(  \mathcal{G}
_{11}^{ss^{\prime}}\left(  \mathbf{x},\mathbf{x}^{+}\right)  \mathcal{+G}
_{22}^{ss^{\prime}}\left(  \mathbf{x},\mathbf{x}^{-}\right)  \right)  \right]
\label{ll}
\end{equation}
where
\begin{equation}
\mathcal{G}_{11}^{ss^{\prime}}\left(  \mathbf{x},\mathbf{x}^{+}\right)
=\left.  \mathcal{G}_{11}^{ss^{\prime}}\left(  \mathbf{x},\mathbf{y}\right)
\right|  _{y\rightarrow x^{+}}.
\end{equation}
The last term is simply the spin of the current. \ Recalling 
that at equal times and equal positions, all different Green's functions are 
related, the equation of motion for $\mathbf{M}$ simply 
becomes
\begin{equation}
\partial_{t}\mathbf{M}\left(  \mathbf{x}\right)  =\mathbf{M}\left(
\mathbf{x}\right)  \times\left[  \frac{1}{2} J \; \nabla^{2}\mathbf{M}
\left(
\mathbf{x}\right)  {+} \; \mathbf{B} + {\lambda} \;
 \mathbf{\mathfrak{M}}\left(  \mathbf{x}\right)  \right].\label{llg}
\end{equation}
 The last term gives 
rise to
dissipation and a contribution to the  precession 
 for magnetic 
multilayers. \cite{Zhang}          \  As we
will see below, this term becomes  $J$-dependent 
in the non-uniform case. \ This latter equation, Eq. (\ref{llg}), is 
the equivalent of the Landau-Lifshitz equation (LL) for the magnetization
in the presence of a current. \ This form is still valid even in the 
presence of an electric field. 
The solution of Eq. (\ref{eq2}) can be represented in 
terms of Feynman
diagrams, Fig. \ref{green}. First, we  define the propagator 
of a non-interacting electron in an external magnetic field $\mathbf{B}$ and 
zero electric field,
\begin{equation}
\mathcal{G}_{ss^{\prime}}^{(0) \;\alpha \beta} \left( \mathbf{x}, \mathbf{y} \right)
=\left\{  \;\left(  i\partial_{{t}_{y}}-\in_{\alpha}\right)
+\frac{1}{2}{\sigma_{ss^{\prime}}^{i}} {B}^{i}\;\right\}
^{-1}\delta_{ss^{\prime}}^{\alpha \beta} \left( \mathbf{x}, \mathbf{y}\right) .
\end{equation}
and expand $\mathcal{G}$ in powers of $\lambda$ using 
Eq. (\ref{eq2}).  \ Keeping only  
terms up to
order $\lambda^{2}$, we have 
\begin{equation}
\mathbf{\sigma} \mathcal{G}=\mathbf{\sigma}\mathcal{G}^{(0)}-
\lambda\sigma\mathcal{G}^{(0)}\mathbf{M}\cdot
\sigma \mathcal{G}^{(0)}
+\frac{1}{2}\lambda^{2}\sigma\mathcal{G}^{(0)}\mathbf{M}\cdot \sigma
\mathcal{G}^{(0)}
\mathbf{M}\cdot \sigma\mathcal{G}^{(0)}+\lambda^{2}
\mathcal{G}^{(0)}
 \sigma \cdot \mathcal{M}\cdot\sigma\mathcal{G}
^{(0)}\sigma\mathcal{G}^{(0)}...\label{g}
\end{equation}
This is a matrix equation and hence integrations over time, 
space and spin degrees of freedom are implicit in 
the above
notation. \ Recently Mills calculated the damping contribution
to order $\lambda$.\cite{mills}$^{,}$ \cite{rebei4} One of his conclusions is that 
this contribution is dependent on the symmetry of the system in 
this case. \ This follows from the fact that the spin propagator 
of the conduction electrons in a ferromagnet is not $O(3)$-invariant,
to first order in $\lambda$ and higher since 
it depends explicitly on $\mathbf{M}$.

Equation (\ref{eq3}) gives the dependence of correlations 
on the
exchange interaction and on the s-d interaction between the 
current and
the medium. Our assumption is that exchange interactions are 
much stronger
than the spin-spin interaction. Hence to lowest 
order, we neglect the latter in 
the equation for the fluctuations. To understand the
meaning of such an equation, we study the 
case with strong exchange interactions, i.e.,
 we take the average magnetization 
to be a
constant and assume the external $\mathbf{B}$ field 
to be small. In this case Eq.
(\ref{eq3}) becomes
\begin{equation}
\partial_{t_{y}}\mathcal{M}^{\times k}_{\alpha\beta}
\left(  \mathbf{y}, \mathbf{z} \right) 
 -\left[
\mathbf{M}\cdot\partial_{t_{y}}\mathcal{M}^{\times k}_{\alpha\beta}
\left(  \mathbf{y}
,\mathbf{z}\right)  \right]  \mathbf{M} = i\epsilon^{\alpha\beta}
 \left(  \mathbf{n}\times
\mathbf{M}\right)  +\int d\mathbf{x}\;\left[ J\left(  \mathbf{x} - 
\mathbf{y}
 \right)  \mathcal{M}^{ \times k}_{\alpha\beta}\left(
\mathbf{x},\mathbf{z}\right)  \times\mathbf{M}\right],\label{fluc}
\end{equation}
where for each $k$, the unit vector $\mathbf{n}$ has components 
$n^{i}=\delta^{ik}$. \ The notation $\mathcal{M}^{\times k}$ is 
for a  vector with components $\mathcal{M}^{i k}, \;  i = 1,2,3$. 
Now if we average over the variable $\mathbf{z}$, we get an 
equation that
gives the time variation of the fluctuations of the magnetization 
around
$\mathbf{M}$. These fluctuations will in turn cause fluctuations 
in the
current through the last term in Eq. (\ref{g}). 
\ Next we show 
how this latter 
equation gives rise to a Boltzmann-type equation for the 
magnetization fluctuations $\mathcal{M}^{i l}$.  
First we expand $\hspace{0pt}\mathcal{M}^{\times k}\left(  \mathbf{x}%
,\mathbf{z}\right)  $ around the 
position $\overrightarrow{\mathbf{y}}$,
\begin{equation}
\mathcal{M}^{\times k}\left(  \mathbf{x},\mathbf{z}\right)  =\mathcal{M}%
^{\times k}\left(  \mathbf{y},\mathbf{z}\right)  +\left.  \partial
_{x}\mathcal{M}^{\times k}\left(  \mathbf{x},\mathbf{z}\right)  \right|
_{\mathbf{x}=\mathbf{y}}\Delta\mathbf{x+}\left.  \frac{1}{2}\partial
_{\mathbf{x}}\partial_{\mathbf{x}}\mathcal{M}^{\times k}\left(  \mathbf{x}%
,\mathbf{z}\right)  \right|  _{\mathbf{x}=\mathbf{y}}\Delta\mathbf{x}%
\Delta\mathbf{x+...}
\end{equation}
where $\Delta\mathbf{x}=\overrightarrow{\mathbf{x}}- 
\overrightarrow{\mathbf{y}}$. If we put this back in Eq. $\left(
{\ref{fluc}}\right)$, we get a diffusion-type equation for all components of the
magnetization fluctuations
\begin{align}
\partial_{t_{y}}\mathcal{M}_{\alpha\beta}^{\times k}\left(  \mathbf{y}
,\mathbf{z}\right)  \mathbf{-}\left[  \mathbf{M}\cdot\partial_{t_{y}%
}\mathcal{M}_{\alpha\beta}^{\times k}\left(  \mathbf{y},\mathbf{z}\right)
\right]  \mathbf{M}+\frac{1}{2}J_{2}\left(  \mathbf{y}\right)
\mathbf{M}\times\nabla^{2}_{y}\mathcal{M}_{\alpha\beta}^{\times
k}\left(  \mathbf{y},\mathbf{z}\right)    & =i\epsilon^{\alpha\beta}%
\mathbf{n}\times\mathbf{M} \label{diff}\\
& -J_{0}\left(  y\right)  \mathbf{M}\times\mathcal{M}_{\alpha\beta}^{\times
k}\left(  \mathbf{y},\mathbf{z}\right)  ,\nonumber
\end{align}
where $J_{0}\left(  \mathbf{y}\right)  $ and $J_{2}\left(  \mathbf{y}\right)
$ are the zeroth and second moment of the exchange coupling,%
\begin{align}
J_{0}\left(  \mathbf{y}\right)    & =\int d\mathbf{x}J\left(  \mathbf{x}%
-\mathbf{y}\right)  ,\\
J_{2}\left(  \mathbf{y}\right)    & = \frac{1}{3}\int d\mathbf{x}J\left(
\mathbf{x}-\mathbf{y}\right)  \Delta x^{2}.
\end{align}
These integrals converge since the exchange 
coupling is short ranged.
\hspace{0pt}The first moment vanishes since we are assuming 
isotropic exchange coupling. \ Hence  treatment of the 
coupling of the conduction electrons to the ambient 
magnetization at low temperatures or temperatures close to $T_{c}$ 
must include Eq.($\ref{diff}$) to account for the fluctuations 
of the magnetization. \

\section{Applications}

\begin{figure}
[ptb]
\begin{center}
\includegraphics[
natheight=8.7027700in,
natwidth=6.127700in,
height=4.0277in,
width=4.0277in
]
{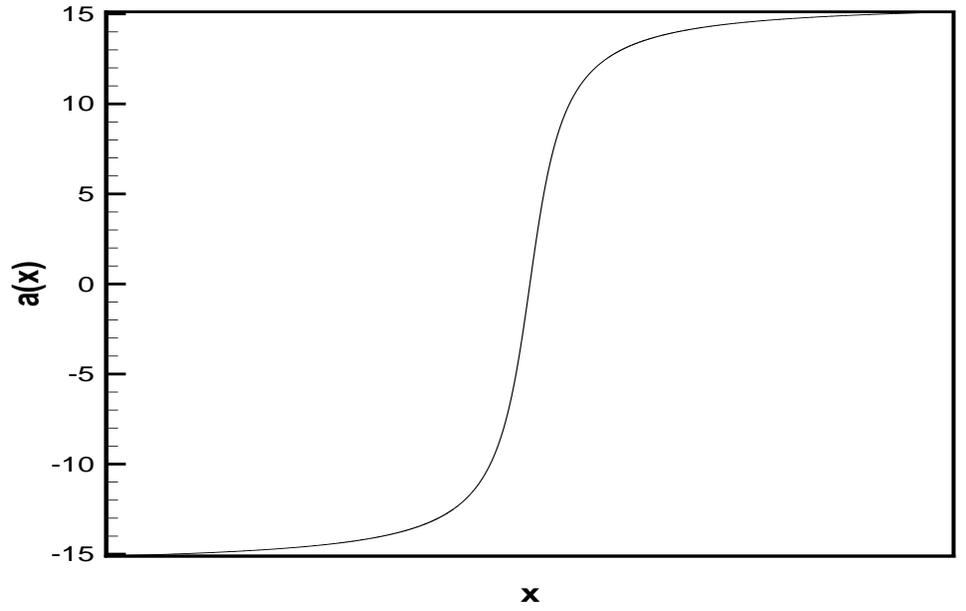}

\caption{Profile of the interface (or molecular field $a(x)$)
 used in the text. The current 
flows perpendicular to the interface. The nonmagnetic 
spacer is taken to have zero thickness.}
\label{interface}
\end{center}
\end{figure}

In this section, we mainly show how this formalism can be applied to 
multilayers. \ In Ref. \onlinecite{hcr}, we showed how our results 
extend those of Zhang, Levy and Fert \cite{Zhang}
 by taking into account the 
indirect exchange effect of the magnetization on the conduction 
electrons. \ This is an important effect in transition metals 
and can not be treated by a Born approximation. \ In the following 
we study two types of structures with nonhomogeneous 
magnetization. \ First we examine
    CPP-type structures with very thin paramagnetic
spacers and no interfacial scattering. \ Second we consider
structures which are 
topologically equivalent to a torus. \ These examples 
clearly illustrate the origin of spin accumulation to be 
directly related to inhomogeneities in the magnetization. \ It is also
obvious from these examples that domain walls are 
another physical example where the results presented here can be 
applied. \ The interface will not be represented 
by a step function in  the 
examples below and will instead take the shape shown
 in Fig. \ref{interface} which plots the mean field $a(x)$ due 
to the magnetization.

 \ For a local  magnetization which is a function only 
of distance x in the direction of the
current, $\mathbf{M}=M\left( x\right) \mathbf{z}$, the 
spin accumulation obeys the simplified equations

\begin{align}
D_{xx}\frac{d^{2}m_{x}\left( x\right) }{dx^{2}}+D_{xy}\frac{d^{2}m_{y}\left(
x\right) }{dx^{2}} &  \notag \\
-2\frac{D_{xy}^{2}}{Da\left( x\right) }\frac{da\left( x\right) }{dx}\frac{%
dm_{x}\left( x\right) }{dx}+\left( D_{xx}-2\frac{D_{xy}^{2}}{D}\right) \frac{%
da\left( x\right) }{dx}\frac{dm_{y}\left( x\right) }{dx} & =\frac{
m_{x}\left( x\right) }{\tau_{sf}}-\frac{a\left( x\right) m_{y}\left( x\right) }{%
\tau_{sf}},
\end{align}
\begin{align}
-D_{xy}\frac{d^{2}m_{x}\left( x\right) }{dx^{2}}+D_{yy}\frac{d^{2}m_{y}\left(
x\right) }{dx^{2}} &  \notag \\
-\left( D_{yy}-2\frac{D_{xy}^{2}}{D}\right) \frac{da\left( x\right) }{dx}%
\frac{dm_{x}\left( x\right) }{dx}-2\frac{D_{xy}^{2}}{Da\left( x\right) }%
\frac{da\left( x\right) }{dx}\frac{dm_{y}\left( x\right) }{dx} & =\frac{%
m_{y}\left( x\right) }{\tau_{sf}}+\frac{a\left( x\right) m_{x}\left( x\right) }{%
\tau_{sf}},
\end{align}
\begin{equation}
D\frac{d^{2}m_{z}\left( x\right) }{dx^{2}}=
\frac{m_{z}}{\tau_{sf}},
\end{equation}
where
\begin{equation}
a\left( x\right) =\tau \lambda M\left( x\right).
\end{equation}
The  coefficients $D_{xx}$, $D_{yy}$, $D_{xy}$ and $D_{yx}$
are functions of the local magnetization, the scattering rates and 
the exchange constant: 
\begin{equation}
D_{p}=D_{xx}=D_{yy}=\frac{D}{1+\left( \tau 
\lambda M(x)\right) ^{2}},
\end{equation}
\begin{equation}
D_{yx}=-D_{xy}=\frac{\tau \lambda M(x)}{1+\left( \tau 
\lambda M(x)\right) ^{2}}.
\end{equation}
 These equations will be solved for different configurations of the 
magnetization $\mathbf{M}$. \ We adopt the following 
parameters for our calculations: \ The spin diffusion length 
$l_{sdl} = \sqrt{D \tau_{sf}} = 100\, nm$, $D = 10^{-3} \;
  m^2/sec$, $D_{yx} = 100\, D_{xx}$, and 
$\lambda = 0.1 \, eV$. \ It should be noted that in these 
equations, the torque term has the opposite sign to that 
which  appears 
in e.g., Zhang et al. \cite{Zhang} since we have taken the electron 
charge to be positive in our definitions of the magnetic 
moments.

\begin{figure}
[ptb]
\begin{center}
\includegraphics[
natheight=8.7027700in,
natwidth=6.127700in,
height=4.0277in,
width=4.0277in
]
{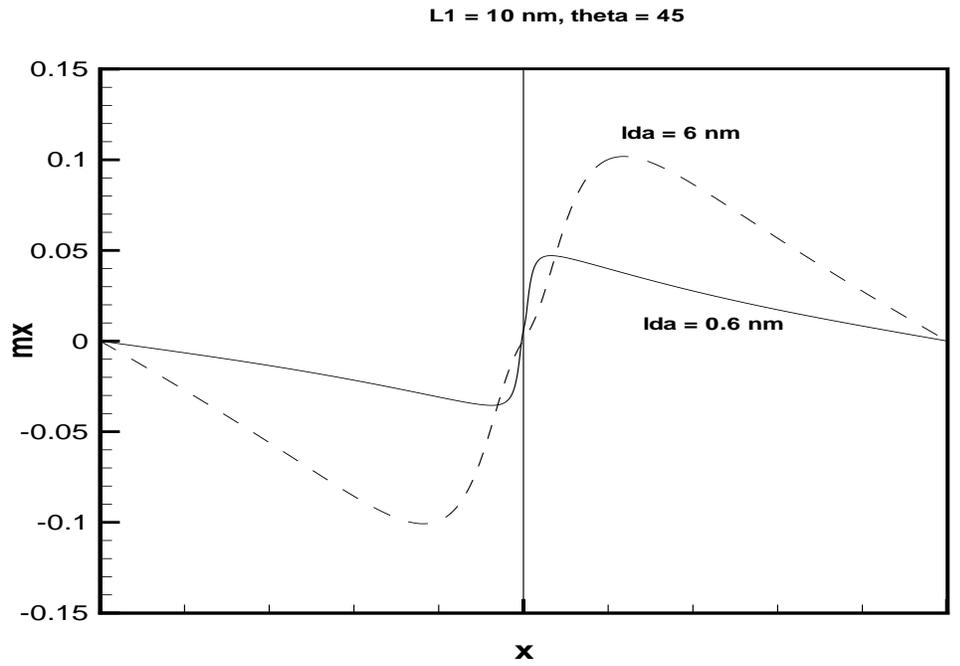}

\caption{The x-component of the 
spin accumulation as a function of the interface
inhomogeneities. On the left, $M = - M_0 \mathbf{z}$ and 
on the right $M = M_0 \mathbf{z}$. \ $lda=0.6 nm$ corresponds to 
the sharper interface.}
\label{thickness_A}
\end{center}
\end{figure}

\begin{figure}
[ptb]
\begin{center}
\includegraphics[
natheight=8.7027700in,
natwidth=6.127700in,
height=4.0277in,
width=4.0277in
]
{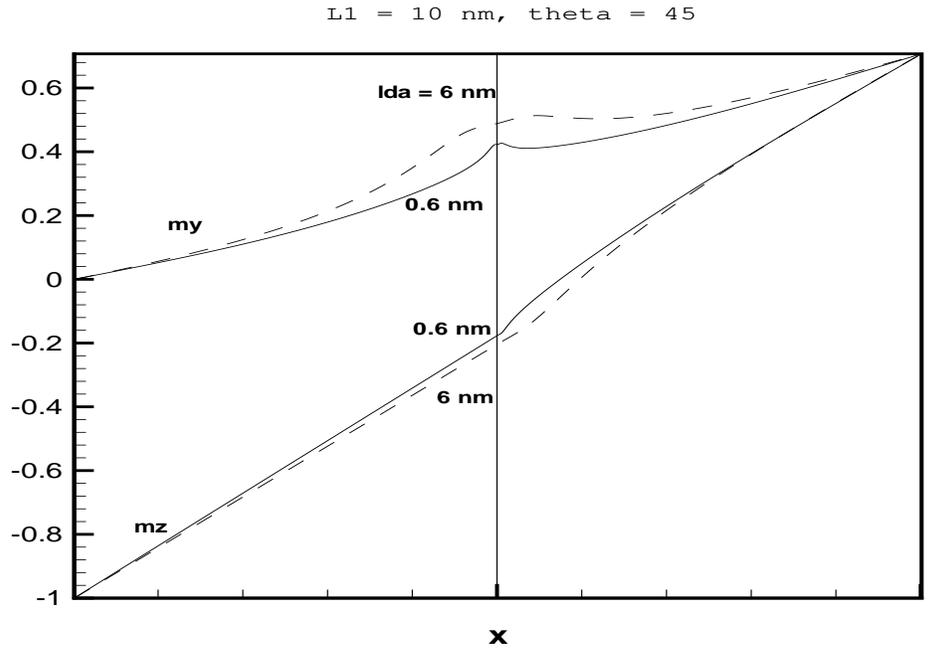}

\caption{The y and z components of the 
spin accumulation as a function of the interface
inhomogeneities for the same configuration of $\mathbf{M}$ as in
 Fig. \ref{thickness_A}.}
\label{thickness_B}
\end{center}
\end{figure}

\begin{figure}
[ptb]
\begin{center}
\includegraphics[
natheight=8.7027700in,
natwidth=6.127700in,
height=4.0277in,
width=4.0277in
]
{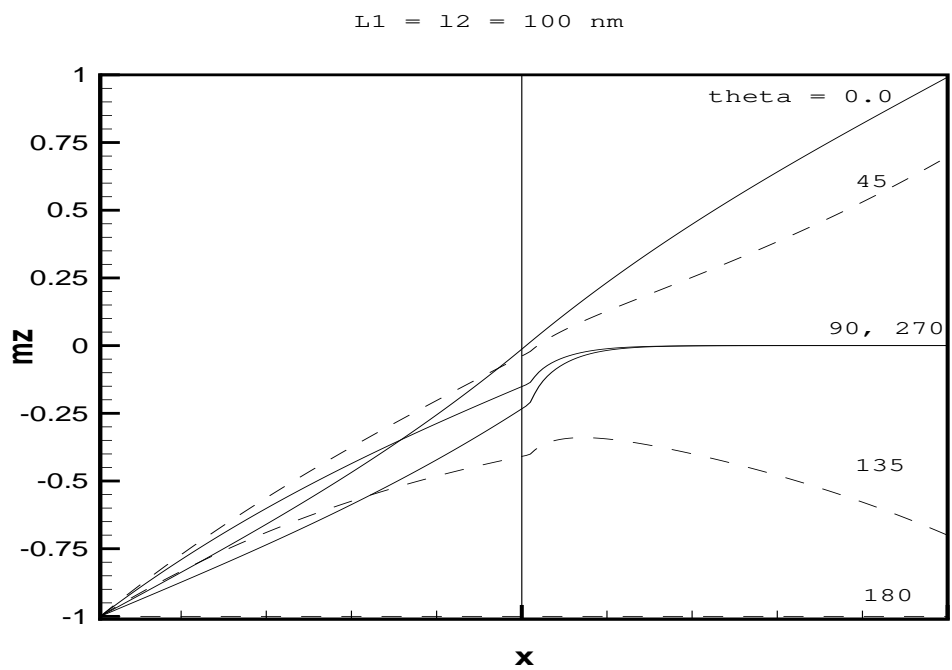}

\caption{The z-component of the spin accumulation as a 
function of the angle for $L = 100 \, nm$. The angles shown on the right 
are expressed in degrees. }
\label{l0100-mz}
\end{center}
\end{figure}

\begin{figure}
[ptb]
\begin{center}
\includegraphics[
natheight=8.7027700in,
natwidth=6.127700in,
height=4.0277in,
width=4.0277in
]
{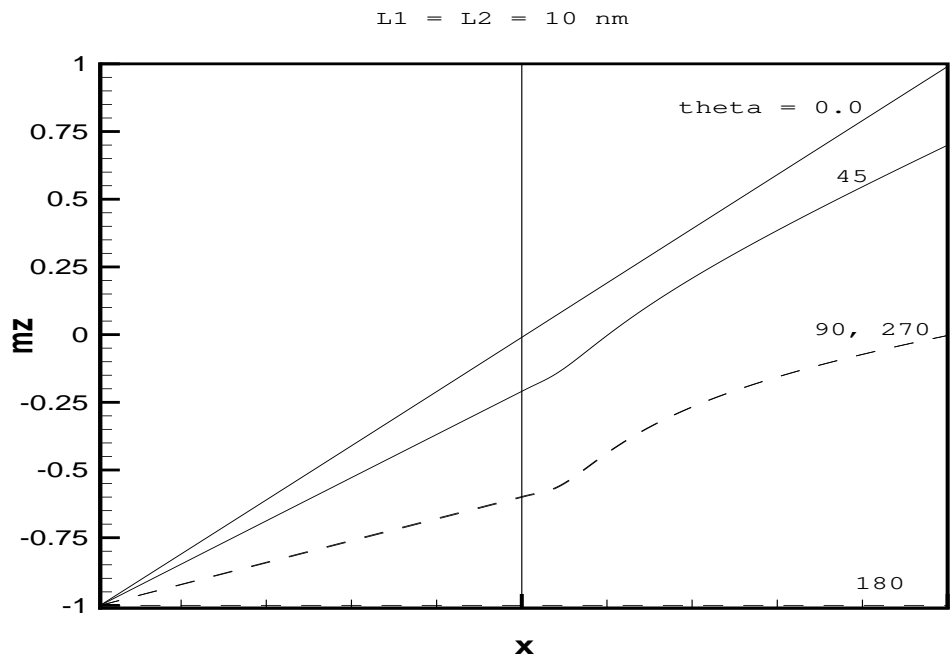}

\caption{Spin accumulation as a function of the angle 
between the magnetization in both regions: z-component for $ L = 10\,
 nm$. The angles are in degrees as in fig. 8.}
\label{l0010-mz}
\end{center}
\end{figure}

\begin{figure}
[ptb]
\begin{center}
\includegraphics[
natheight=8.7027700in,
natwidth=6.127700in,
height=4.0277in,
width=4.0277in
]
{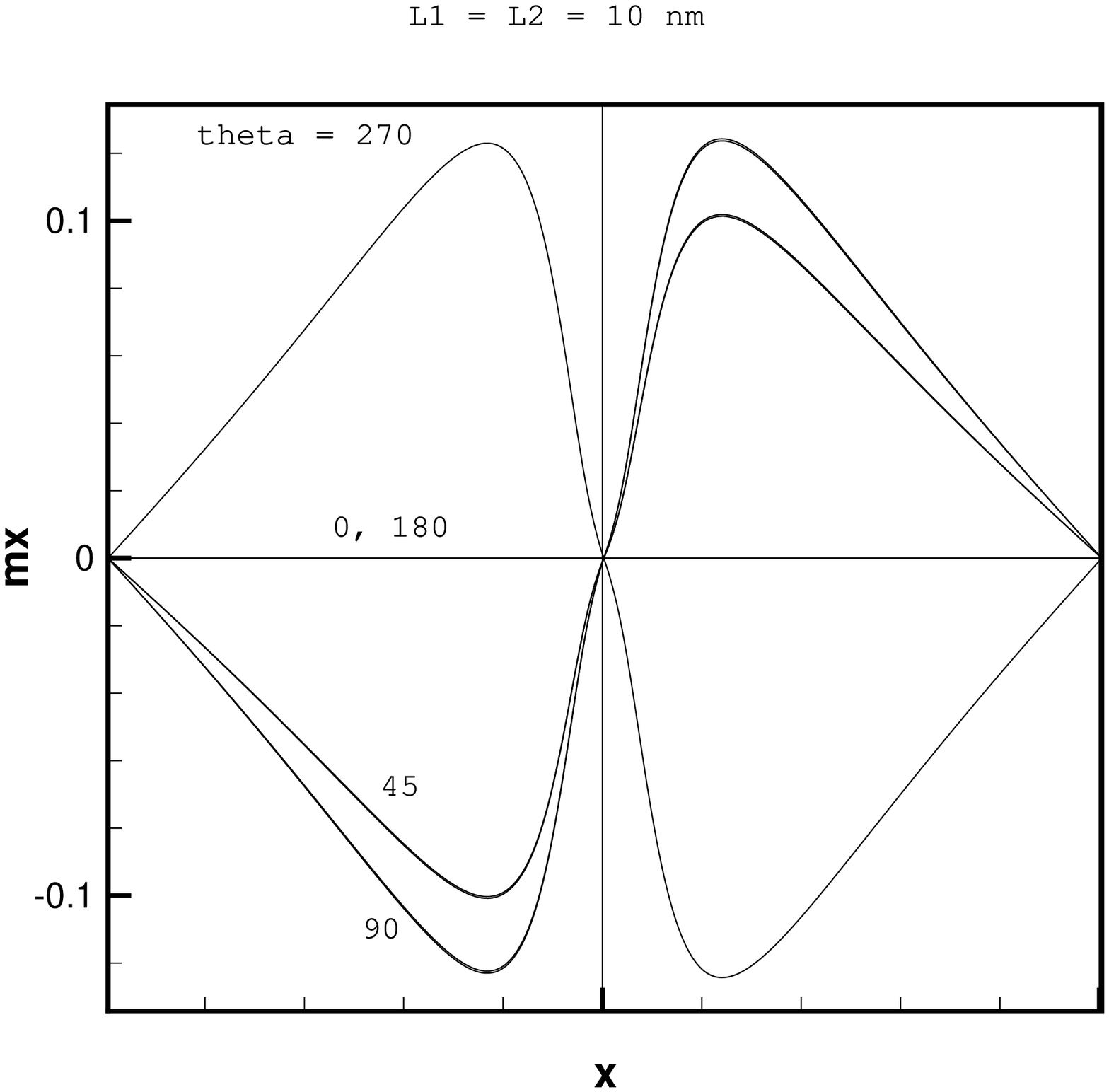}

\caption{x-component of the spin accumulation 
as a function of the angle 
between the magnetization in both regions for $L= 10 \, nm$.}
\label{l0010-mx}
\end{center}
\end{figure}

\begin{figure}
[ptb]
\begin{center}
\includegraphics[
natheight=8.7027700in,
natwidth=6.127700in,
height=4.0277in,
width=4.0277in
]
{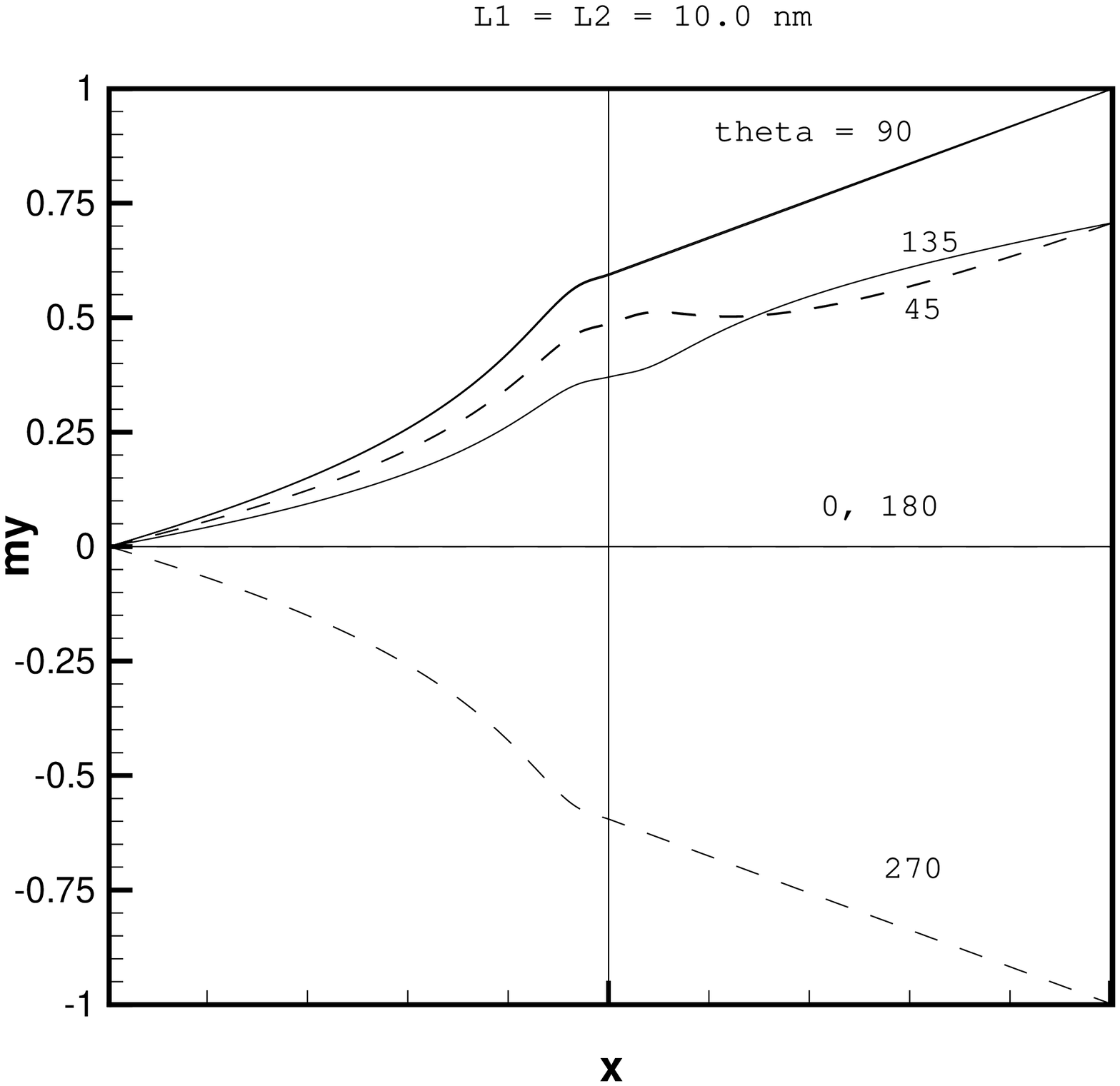}

\caption{y-component of spin accumulation as a function of the angle 
between the magnetization in both regions for $L = 10 \, nm$.}
\label{l0010-my}
\end{center}
\end{figure}

\begin{figure}
[ptb]
\begin{center}
\includegraphics[
natheight=8.7027700in,
natwidth=6.127700in,
height=4.0277in,
width=4.0277in
]
{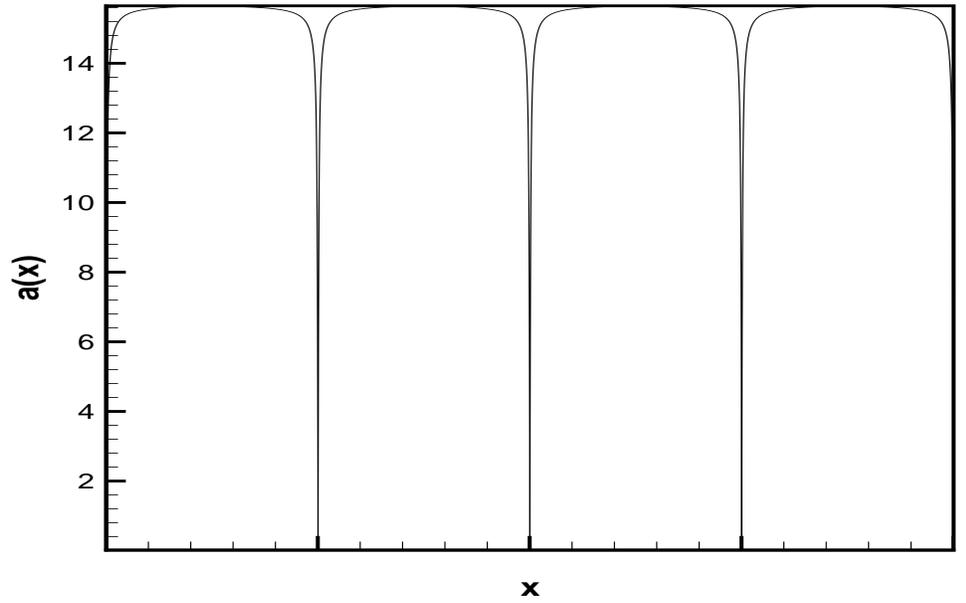}

\caption{Profile of the local magnetization along the  
different sides of the square ring (or torus) in one period. The 
magnetization is in the yz-plane whereas the current is in the 
x direction. \ The relative angle 
of the magnetization between neighboring sides is 90 degrees.}
\label{4s-a}
\end{center}
\end{figure}

\begin{figure}
[ptb]
\begin{center}
\includegraphics[
natheight=8.7027700in,
natwidth=6.127700in,
height=4.0277in,
width=4.0277in
]
{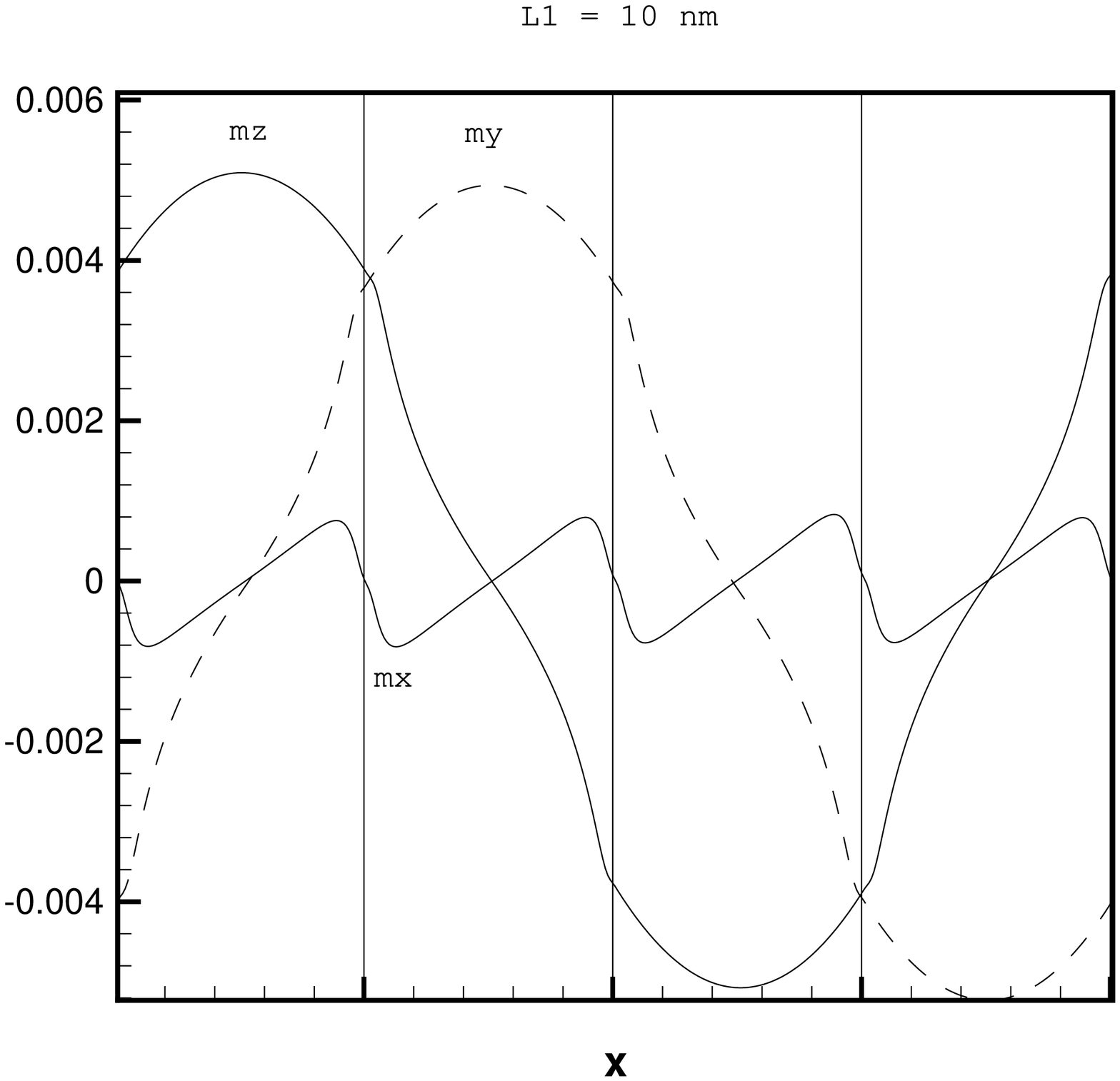}

\caption{Spin accumulation in a torus strucure with 
equal sides intersecting at right angles. The length of each side is 
 L = 10 nm}
\label{l10}
\end{center}
\end{figure}

\begin{figure}
[ptb]
\begin{center}
\includegraphics[
natheight=8.7027700in,
natwidth=6.127700in,
height=4.0277in,
width=4.0277in
]
{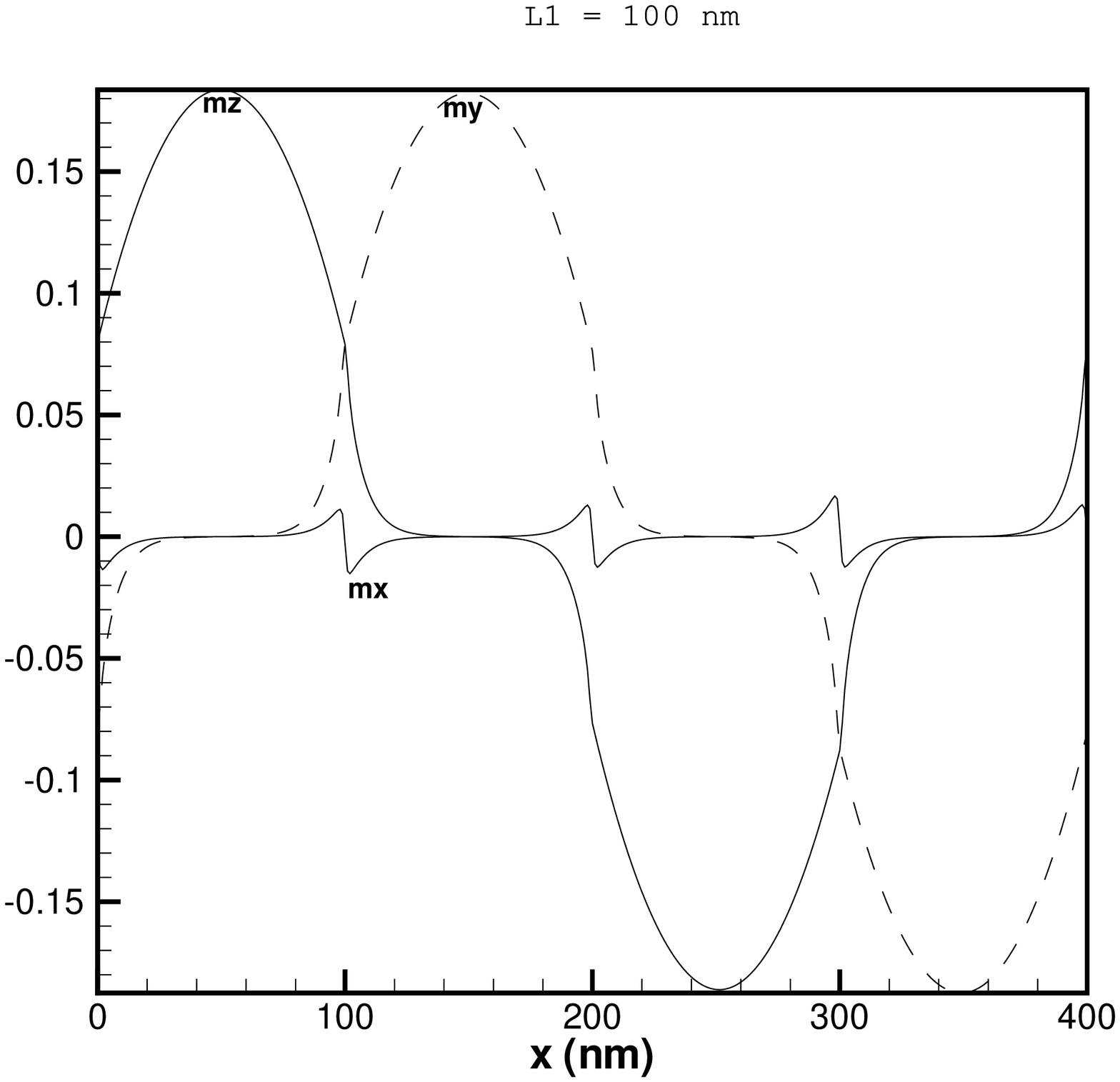}

\caption{Spin accumulation in a torus strucure with 
equal sides intersecting at right angles. The length of each side is   L = 100 nm}
\label{l100}
\end{center}
\end{figure}

\begin{figure}
[ptb]
\begin{center}
\includegraphics[
natheight=8.7027700in,
natwidth=6.127700in,
height=4.0277in,
width=4.0277in
]
{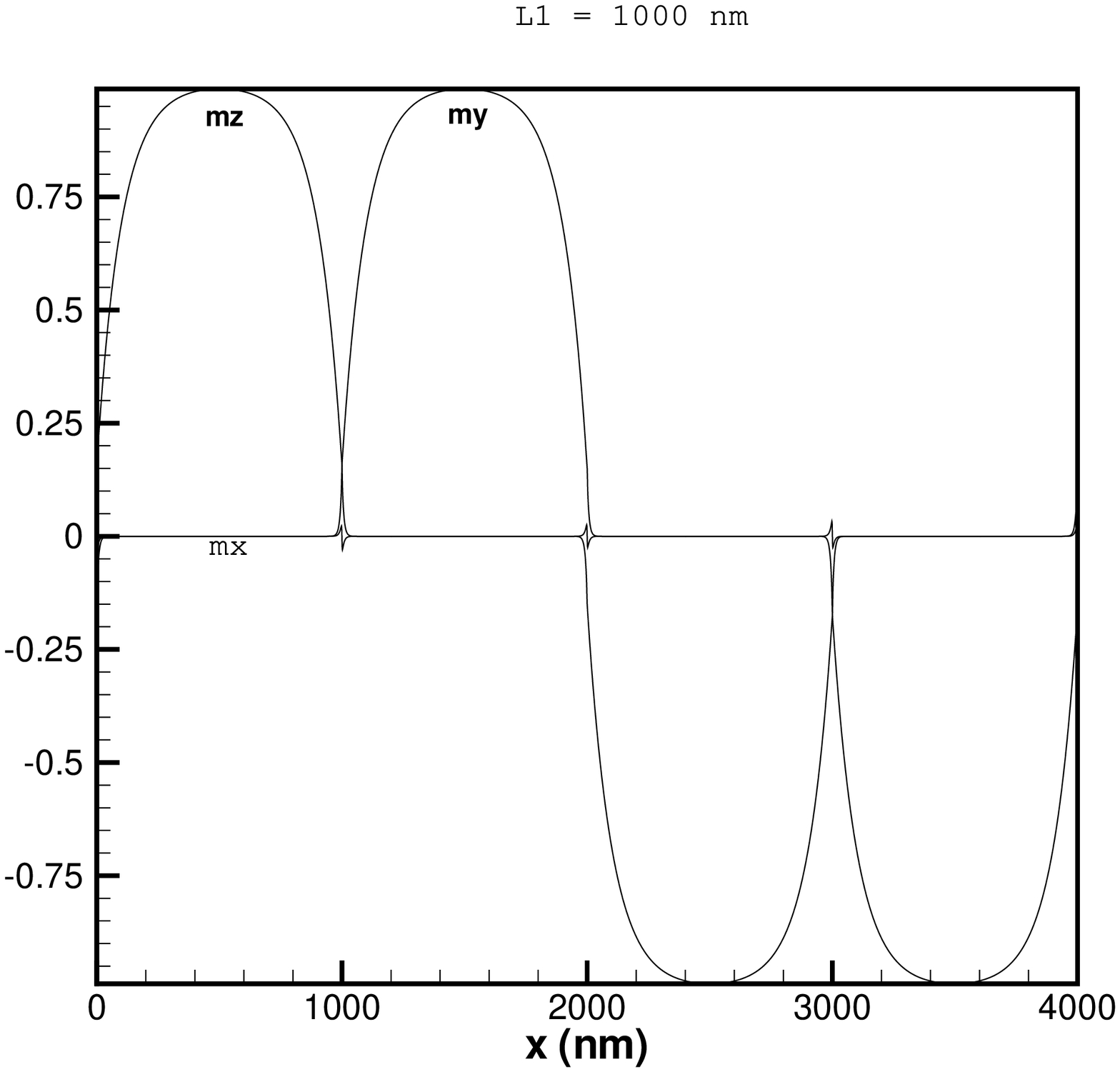}

\caption{Spin accumulation in a torus strucure with 
equal sides intersecting at right angles. The length of each side is 
  L = 1000 nm}
\label{l1000}
\end{center}
\end{figure}

\bigskip

\ First we consider a configuration with in-plane magnetization. \ The 
magnetization is assumed to vary with position along the 
direction of the current. \ The spacer has practically zero 
thickness,
 which is a reasonable approximation for most GMR devices. \ Fig. 
\ref{interface} shows $a(x)$ for a
 typical interface. \ We do not explicitly include a non-magnetic
spacer but we set 
the magnetization  to
zero at the center. \ At the ends it is parallel to the local z-axis, 
the local direction of the equilibrium magnetization.  \ The 
transverse components 
of the spin accumulation are set to zero at the outer ends. 
\ The respective z-axes make a non-zero angle  in order for 
the spin 
accumulation to be non-zero. \ First we demonstrate the effect of 
inhomogeneities on the spin accumulation; we keep the relative 
angle between the magnetizations the same but we vary the size  
of the 'domain wall', the transition region 
of the local mean field. \ It is seen from Fig. \ref{thickness_A} and 
Fig. \ref{thickness_B} that 
the larger the inhomogeneities the larger is the spin accumulation. \ 
This effect is independent of the relative orientations and 
was not predicted before. \ Therefore 
spin accumulation can be 
enhanced by having a layer with constant direction magnetization but 
with spatial inhomogeneities. \ Such a structure can be achieved 
by e.g. having a temperature
 gradient across the slab or controlled 
doping  that changes the magnetic saturation along the direction of 
the current. \ Figures \ref{l0100-mz} and \ref{l0010-mz} show 
the variation of the 
z-component of the spin accumulation with respect to the relative 
angle of the magnetizations and size of the sample. \ Figures 
\ref{l0010-mx} and \ref{l0010-my} 
are for the x and y components for the smaller sample. \ Here, the spin accumulation is largest
for the case where the two magnetizations are orthogonal to each other and the size 
of the sample is smallest. \ In all these results, the equilibrium spin accumulation is normalized to -1 
on the left hand side and normalized to +1 on the right hand side. \ The components of 
the spin accumulation are taken with respect to a global frame,
 that of the layer 
on the left.

\ As our second example, we choose a `ring' structure. \ This can 
be part of a long solenoid with a square cross section. \ Hence 
we now solve our equations  with periodic boundary 
conditions. \ In each side of the square cross section, the 
profile of the magnetization within a 
period  is  shown in fig. \ref{4s-a}. \ The relative angle 
of the magnetization between neighboring sides is $90$ 
degrees. \ We 
study the spin accumulation as a function of the size of the 
sides. \ In figs. \ref{l10}, \ref{l100} and \ref{l1000} we plot 
the three components of the total spin accumulation. \ The 
equilibrium spin accumulation is taken to be normalized to one. \ 
The solutions show the expected behavior. \ The spin 
accumulation tries to reach its equilibrium value near 
the middle of each side. \ The spin accumulation is largest 
when the length of each side is smallest, as expected. \ This 
geometry shows how spin 
accumulation can be transported over large distances and 
also modulated by controlling the size of the cross section, 
similar to what happens in a regular transformer except here 
we are working with spin charge.

\section{Conclusion}
\bigskip We have introduced a 
many-body formalism  based on path-integral 
techniques capable of handling a system of 
 both  local magnetic 
moments  and  conduction electrons  in a self-consistent 
manner. Transport properties can be obtained through the calculation 
of the two-point functions of the 
current and the magnetization respectively. \ One of the important 
outcomes of this treatment is that we were able to derive a set of 
new equations that are needed when the magnetization of 
the medium is no longer homogeneous. \ First we showed that the polarization
of the current is no longer homogeneous and satisfies a generalized 
diffusion equation where the diffusion tensor is dependent on the 
direction of the magnetization. \ We have hence
shown that exchange effects are important 
in a ferromagnet and need to be taken into account properly. 
 \ The fluctuations 
of the magnetization were also shown to obey a  
diffusion-type equation which  depends on 
the direction of the local magnetization. \ This latter equation is 
in addition to 
recovering a Boltzmann equation
for the current which 
follows from Eq. (\ref{current}) and a
 Landau-Lifshitz equation for the average of the 
magnetization, Eq. (\ref{llg}). \  
 We have also shown how the 
non-uniform magnetization affects both the conduction electrons 
 and the 
spin-momentum transfer term. We 
gave a simple physical picture for our main results. \ We 
finally showed how our results can 
be applied in various configurations. \ Our results show that 
spin accumulation can be enhanced by inhomogeneities at the interface. 

\ In this work we focused on spin accumulation and we did not 
deal with its effect on the dynamics of the local 
magnetization. \ We showed how to recover the Landau-Lifshitz 
equation but we discussed the effect of the collisions 
on the classical magnetization only in qualiative 
terms. \ The use of the relaxation time approximation is another 
shortcoming of this work. \ Its improvement will complicate 
the treatment further. \ We believe these issues should be addressed
in the future.

\acknowledgments

We would like to thank 
D. Boyanovski, M. Covington  and M. Simionato for very useful 
discussions. \ After this work 
has been completed, we learned of references  \onlinecite{hirst}
and \onlinecite{kaplan} from  
L. Berger.

\newpage

\newpage

\bigskip

\begin{thebibliography}{99}
\bibitem{Berger}L. Berger, Phys. Rev.B \textbf{54}, 9353 (1996), J. Apl. Phys.
\textbf{89}, 5521 (2001).

\bibitem{Slon}J. C. Slonczewski, J. Magn. Magn. Mater. 
\textbf{159}, L1 (1996).

\bibitem{Cornell}J. A. Katine, F. J. Albert, R. A. Buhrman, E. B. Myers, and D. C.
Ralph, Phys. Rev. Lett.\textbf{84}, 3149 (2000).

\bibitem{Zurich}W. Weber, S. Rieseen, and H. C. Siegmann, Science 
\textbf{291}, 1015 (2001).

\bibitem{Heidi}C. Heidi, P. E. Zilberman and R. J. Elliott, Phys. Rev. B \textbf{63},
64424 (2001). C. Heidi, Phys. Rev. Lett. \textbf{87}, 197201 (2001).

\bibitem{Waintal}X. Waintal, E. B. Myers, P. W. Brouwer, and D. C. Ralph, Phys.
Rev. B \textbf{62}, 12317 (2000).

\bibitem{Zhang}S. Zhang, P. Levy, A. Fert, Phys. Rev. Lett. \textbf{88}, 236601 (2002).

\bibitem{Stiles} M. D. Stiles, A. Zangwill, Phys. Rev. B \textbf{66},
014407, (2002).

\bibitem{Langreth}D. C. Langreth and J. W. Wilkins, Phys. Rev. B \textbf{6}, 3189 (1972).


\bibitem{Chou}K. Chou, Z. Su, and L. Yu, Phys. Rep.\textbf{118}, 1 (1985).


\bibitem{hcr} W. N. G. Hitchon, R. W. Chantrell and A. Rebei, (submitted 
to Phys. Rev. B).


\bibitem{rebei4} A. Rebei, M. Simionato, (to be published).



\bibitem{Schwinger}J. Schwinger, J. Math. Phys. \textbf{2}, 407 (1961).

\bibitem{rebei5} A. Rebei, (unpublished).

\bibitem{Calzetta}E. Calzetta and B. L. Hu, Phys. Rev.D \textbf{37}, 2878 (1988).

\bibitem{Manson}M. Manson and A. Sjolander, Phys. Rev.B \textbf{11}, 4639 (1975).

\bibitem{Feynman}R. P. Feynman and F. L. Vernon, Ann. Phys. (N.Y.) \textbf{24}, 118 (1963).

\bibitem{Rebei}A. Rebei and G.J. Parker, Phys. Rev. B \textbf{67}, 104434 (2003).


\bibitem{rebei2}A. Rebei, M. Simionato and G. J. Parker, Phys. Rev. B 
\textbf{69}, 134412-1 (2004).

\bibitem{Perm}A. Perelemov: Generalized Coherent States and their
Applications, Springer-Verlag, Berlin, 1985.

\bibitem{Itzykson}C. Itzykson and J-B. Zuber, Quantum Field Theory,
McGraw-Hill, New York, 1980.


\bibitem{fradkin} E. Fradkin, Field Theories of Condensed Matter 
Systems, Addison Wesly, New York, 1991.

\bibitem{Bazaliy}Ya. B. Bazaliy, B. A. Jones, and S.-C. Zhang, Phys. Rev.B \textbf{57},
R3213 (1998).

\bibitem{Ferry}H. Haug and A.-P. Jauho, Quantum Kinetics in 
Transport 
and Optics of Semiconductors, Springer, Berlin, 1996.

\bibitem{Baym}L. P. Kadanoff and G. Baym, 
Quantum Statistical Mechanics,
Benjamin, New-York, 1962.

\bibitem{Jackiw}J. M. Cornwall, R. Jackiw, and E. Tomboulis, 
Phys. Rev. D \textbf{10},
2428 (1974).

\bibitem{rebei3} A. Rebei and W. N. G. Hitchon, Int. J. Mod. Phys. B 
\textbf{17}, 973 (2003).

\bibitem{davies} R. W. Davies and F. A. Blum, Phys. Rev. B \textbf{3}, 3321 
(1971).

\bibitem{rammer} J. Rammer and H. Smith, Rev. Mod. Phys. \textbf{58}, 323 (1986). 



\bibitem{bittencourt}J. A. Bittencourt, Fundamentals of Plasma Physics,
Pergamon Press, Oxford 1986.


\bibitem{hirst} L. L. Hirst, Phys. Rev. \textbf{141}, 503 (1966).

\bibitem{kaplan}J. I. Kaplan, Phys. Rev. \textbf{143}, 351 (1966).

\bibitem{mills} D. L. Mills, Phys. Rev. B \textbf{68}, 014419 (2003).





\end{thebibliography}
\end{document}